\documentclass[pra,reprint,amsfonts,amsmath,amssymb,showpacs,twocolumn]{revtex4}

\usepackage[dvipdfmx]{graphicx}
\usepackage{dcolumn}
\usepackage{bm}
\usepackage{amsthm}
\usepackage{mathrsfs}
\usepackage{comment}
\usepackage{color}
\begin{document}
\title{Topological classification of vortex-core structures of spin-1 Bose-Einstein condensates}

\author{Shingo Kobayashi$^{1,2}$}
\author{Yuki Kawaguchi$^1$}
\author{Muneto Nitta$^2$}
\author{Masahito Ueda$^1$}
\affiliation{$^1$Department of Physics, University of Tokyo, 7-3-1, Hongo, Bunkyo-ku, Tokyo 113-0033, Japan  \\ $^2$Department of Physics, and Research and Education Center for Natural Sciences, Keio University, 4-1-1 Hiyoshi, Yokohama, Kanagawa 223-8511, Japan}
\date{\today}
\begin{abstract}
We classify vortex-core structures according to the topology of the order parameter space. We develop a method to characterize how the order parameter changes inside the vortex core. We apply it to spin-1 Bose-Einstein condensates and show that the vortex-core structures are classified by winding numbers that are locally defined in the core region. We also show that a vortex-core structure with a nontrivial winding number can be stabilized under a negative quadratic Zeeman effect. 
\end{abstract}
\maketitle
\section{Introduction}
\label{sec:intro}
One of the salient features of superfluidity is the quantization of vortices. Due to the single-valuedness of the macroscopic wavefunction, the circulation of a scalar superfluid is quantized in units of $h/M$, where $h$ is the Planck constant and $M$ is the mass of the constituent particles. 
However, the situation changes in a very significant manner when the system has internal degrees of freedom. 

Since spinor Bose-Einstein condensates (BECs), namely, BECs with spin degrees of freedom, have been realized in ultracold atomic systems, a number of studies have been conducted to understand macroscopic quantum phenomena, including experimental~\cite{Matthews:1999,Madison:2000,Abo:2001,Leanhardt:2002,Leanhardt:2003,Sadler:2006} and theoretical~\cite{Dalfovo:1996,Yip:1999,Isoshima:2001,Martikainen:2002,Mizushima:2002,Ruostekoski:2003} studies on quantized vortices. In a spinor BEC, the circulation is quantized in units of a rational fraction of $h/M$, or not quantized at all. This is because the spin and gauge degrees of freedom are coupled in a nontrivial manner. Examples include half-quantum vortices in the spin-1 polar phase~\cite{Leonhardt;2000} and $1/3$-quantum vortices in the spin-2 cyclic phase~\cite{Semenoff;2007,michikazu;2009,Makela:2003}.

Vortices in spinor BECs have been extensively investigated and classified using homotopy theory~\cite{Mermin;1979,Trebin;1982,Michel;1980,Mineev;1998,Volovik;2003}. However, the conventional homotopy theory does not tell much about the structure of a vortex core. For the case of a scalar BEC, the particle number density should vanish at the vortex core in order to avoid the phase singularity. We call such a vortex, i.e., a vortex accompanied by a density hole, a singular vortex. In contrast, for the case of a spinor BEC, the particle density does not have to vanish at the vortex core. Since the spinor BEC is described with a multi component order parameter, even when one component has a phase singularity, other components can fill the vortex core. As a whole, the particle density may become nonzero. We call this type of vortex a nonsingular vortex. A nonsingular vortex was realized in a spin-1 ferromagnetic BEC~\cite{Sadler:2006}. Recently, vortex-core structures have been numerically investigated for spin-2 BECs~\cite{kobayashi:2009}.           

In this paper, we classify the vortex-core structure from the point of view of topology. Here, we define the vortex core and the core structure as follows. In general, a spontaneously broken-symmetry state is characterized by a coset space $G/H$, where $G$ is the full symmetry of the system and $H$ is a remaining symmetry of the broken-symmetry state (isotropy group). This coset space is called an order parameter manifold (OPM). We denote it as $\mathcal{M}_1 \simeq G/H$ in this article. In general, the classification of vortices is given by calculating the fundamental group of the OPM~\cite{Mermin;1979,Trebin;1982,Michel;1980,Mineev;1998,Volovik;2003}. However, this classification is applicable only to a region which is far away from the vortex because, close to the vortex core, the kinetic energy associated with the circulating current increases and the order parameter goes out of the OPM. We define the {\it vortex core} as the region in which the order parameter leaves the OPM, and the term {\it vortex-core structure} is meant to represent how the order parameter changes in the vortex core region~\cite{text1}.   
A topological method of classifying the vortex core structure was first proposed by Mermin {\it et al}.~\cite{Memin:1978} and  Lyuksyutov~\cite{Lyuksyutov:1978}: this method tells us whether a vortex is singular or nonsingular. However, there remains a question as to what the vortex-core structure is when the vortex is nonsingular. We address this question in the present paper by developing a general method to classify the vortex-core structure of a nonsingular vortex. Our method gives all topologically possible vortex-core structures. We then apply our method to spin-1 BECs and find that the vortex core can accommodate not just one state but many different types of states which are arranged in a concentric fashion. We characterize such a complex structure by introducing a local winding number along the concentric circle. Because the homotopy theory can enumerate possible states but cannot demonstrate their existence, we show that such vortex-core structures can indeed be realized by solving a time-dependent Gross-Pitaevskii equation numerically.  

 The paper is organized as follows. In Sec.~\ref{sec:review}, we review the classification of vortex singularities according to Refs.~\cite{Memin:1978,Lyuksyutov:1978} and apply it to spin-1 BECs. In Sec.~\ref{sec:search}, we generalize the method in order to classify the vortex-core structure. In Sec.~\ref{sec:vortex}, we apply our classification method to spin-1 BECs. We show that vortex cores are made up of concentric patterns in different states. In Sec.~\ref{sec:dyn}, we verify that these vortex-core structures can indeed be realized as energetically stable configurations by numerical simulations under a magnetic field. Finally, in Sec.~\ref{sec:sum}, we summarize our results.

\section{Review of the classification of the vortex core}
\label{sec:review}
A mathematical scheme to classify a state at a vortex core was first proposed by Mermin {\it et al}.~\cite{Memin:1978} and Lyuksyutov~\cite{Lyuksyutov:1978} who applied the scheme to classify vortex-core structures in the superfluid Helium-$3$ and those in a liquid crystal, respectively. In this section, we briefly review their classification method and illustrate the method for the case of a spin-1 BEC. 
\subsection{Classification of the vortex core}
\label{sec:MVM}
We start with the energy functional given by 
\begin{align}
 E = \int  d \bm{r} (\epsilon_{\text{kin}} + \epsilon_{\text{int}}),
\end{align}
where $\epsilon_{\text{kin}}$ is the kinetic energy density and $\epsilon_{\text{int}}$ is the interaction energy density.
Suppose that the interaction energy density can be decomposed into two terms, $\epsilon_1$ and $\epsilon_2$, as
\begin{align}
 \epsilon_{\text{int}} = \epsilon_1 + \epsilon_2, \label{eq:intene}
\end{align}  
and that the energy scales of these terms are quite different, say, $|\epsilon_1| \ll |\epsilon_2|$. The ground state in a uniform system is given by minimizing the interaction energy $\epsilon_{\text{int}}$. Note that when the symmetry of the system is spontaneously broken in the ground state, we cannot uniquely determine the ground state. Instead, we can construct a degenerate space of order parameters that minimize $\epsilon_{\text{int}}$. For example, if the global $U(1)$ gauge symmetry is spontaneously broken, all states obtained by applying $U(1)$ gauge transformations to a ground state also minimize the interaction energy. Such a degenerate space of order parameters is called an OPM, which is defined by
\begin{align}
\mathcal{M}_1 := \{ g \psi | ^{\forall} g \in G/H_{\psi}\} \simeq G/H_{\psi}, \label{eq:defOPM}
\end{align}
where $\psi$ is a ground state of $\epsilon_{\rm int}$. We define a full symmetry group of system $G$, the action of which does not change the interaction energy, and  an {\it isotropy group} of $\psi$, which makes $\psi$ invariant under a group action:
\begin{align}
H_{\psi} = \{ g \in G | g \bm{\psi} = \bm{\psi} \}. \label{eq:defH}
\end{align}
 We also introduce an enlarged OPM $\mathcal{M}_2$ as a space of the order parameters that minimize $\epsilon_2$, where $\mathcal{M}_1$ and $\mathcal{M}_2$ satisfy 
\begin{align}
 \mathcal{M}_1 \subseteq \mathcal{M}_2.\label{eq:m1<m2}
\end{align} 

In the presence of a vortex, the order parameter becomes space-dependent, and the kinetic energy density is roughly given by $\epsilon_{\rm kin} \sim \hbar^2n/(2 M r^2) $, where $n$ is the particle-number density, $M$ the mass of the particle, and $r$ the distance from the vortex core. Hence, the kinetic energy density becomes negligible far from the vortex core and the order parameter in that region belongs to $\mathcal{M}_1$. 
On the other hand, close to the vortex core, the kinetic energy density becomes comparable to the interaction energy density. In the region of $ |\epsilon_1| \lesssim \epsilon_{\text{kin}} \lesssim |\epsilon_2|$, $\epsilon_{\rm kin}$ dominates the weaker interaction energy $\epsilon_1$ and the order parameter belongs no longer to $\mathcal{M}_1$ but to an enlarged OPM $\mathcal{M}_2$.

 According to the homotopy theory~\cite{Mermin;1979,Trebin;1982,Michel;1980,Mineev;1998,Volovik;2003}, a topologically stable vortex is labeled with an element of the fundamental group $\pi_1 (\mathcal{M}_1)$.
Since the OPM near the vortex core is enlarged to $\mathcal{M}_2$, the topological structure near the core is classified with $\pi_1 (\mathcal{M}_2)$. Here, we consider a loop $l$ in the OPM $\mathcal{M}_1$ that cannot continuously shrink to a point in $\mathcal{M}_1$. The homotopy equivalent class of $l$ characterizes a vortex. The loop $l$ is embedded in the enlarged OPM $\mathcal{M}_2$. If the loop $l$ shrinks to a point in $\mathcal{M}_2$, the core is nonsingular because the state in the core is specified by a point in $\mathcal{M}_2$.  On the other hand, if the loop $l$ cannot continuously shrink to a point in $\mathcal{M}_2$,  the corresponding vortex is topologically stable near the vortex core.  Thus, the vortex can shrink to a point when we extend the enlarged OPM $\mathcal{M}_2$ to the entire degrees of freedom of the order parameter, which includes a point at which the order parameter vanishes. Therefore, the vortex core is singular. The above statement can be expressed by using an inclusion map from $\pi_1 (\mathcal{M}_1)$ to $\pi_1 (\mathcal{M}_2)$. Because $\mathcal{M}_1$ is embedded in $\mathcal{M}_2$, there is an inclusion map from $\pi_1 (\mathcal{M}_1)$ to $\pi_1 (\mathcal{M}_2)$:
\begin{align}
 \Phi : \; \pi_1 (\mathcal{M}_1) \to \pi_1 (\mathcal{M}_2). \label{eq:MVM}
\end{align}
For a given $\gamma \in \pi_1 (\mathcal{M}_1)$, if $\Phi(\gamma) = 1_c$, the vortex labeled $\gamma$ is nonsingular, whereas if $\Phi (\gamma) \neq 1_c$, the core becomes singular.  Here, $1_c$ is an identity element of $\pi_1(\mathcal{M}_2)$. The singularity is characterized by an image of $\Phi$. 

\subsection{Application to spin-1 BECs}
\label{sec:applspin1}
We apply the above method to spin-$1$ BECs. The order parameter of a spin-1 BEC is given by
\begin{align}
 \bm{\psi} = (\psi_1, \psi_{0}, \psi_{-1})^T \in \mathbb{C}^3, \label{eq:opspin1}
\end{align}
where $\psi_m \; (m=1,0,-1)$ describes a macroscopic wave function for atoms in the magnetic sublevel $m$.
The mean-field energy functional for a uniform system is given by~\cite{Ohmi:1998,Ho:1998,Ueda:2010}
\begin{align}
 E = \int d \bm{r} \left( \epsilon_{\text{kin}} + \epsilon_{\text{int}} \right), \label{eq:enetot}
\end{align}
where $\epsilon_{\text{kin}}$ and $\epsilon_{\text{int}}$ are respectively given by
\begin{subequations}
\begin{align}
&\epsilon_{\text{kin}} =  \sum_{m=-1}^1 \frac{\hbar^2 }{2M} |\nabla \psi_m|^2 ,\label{eq:enekin}\\
&\epsilon_{\text{int}} = \frac{c_0 n^2}{2} + \frac{c_1 n^2}{2} |\bm{F}|^2. \label{eq:enebulk}
\end{align}
\end{subequations}
Here, the first and second terms in Eq.~(\ref{eq:enebulk}) represent the spin-independent and spin-dependent energy densities, respectively, and $n$ and $|\bm{F}|$ are the local density and local magnetization, given by
\begin{align}
 &n = \sum_{m=-1}^1 |\psi_m|^2, \label{eq:dos}\\
 &|\bm{F}| = \frac{1}{n}\sqrt{\sum_{m,m'=-1}^1 \psi_m (\bm{f})_{mm'} \psi_{m'}} \label{eq:lmag}, 
\end{align}
 respectively, with $\bm{f} = (f_x, f_y ,f_z)$ being a vector of the spin-1 matrices given by
 \begin{align}
  f_x &= \frac{1}{\sqrt{2}} \begin{pmatrix} 0 & 1 & 0 \\ 1 & 0 & 1 \\ 0 & 1 & 0 \end{pmatrix}, \\ \notag f_y &= \frac{i}{\sqrt{2}} \begin{pmatrix} 0 & -1 & 0 \\ 1 & 0 & -1 \\ 0 & 1 & 0 \end{pmatrix}, \\ \notag f_z &= \begin{pmatrix} 1 & 0 & 0 \\ 0 & 0 & 0 \\ 0 & 0& -1 \end{pmatrix}. 
 \end{align} 
 The coefficients of $c_0$ and $c_1$ are given by
 \begin{align}
 c_0 = \frac{4 \pi \hbar^2 }{M} \frac{a_0+ 2 a_2}{3}, \ \  c_1 = \frac{4 \pi \hbar^2 }{M}\frac{a_2 - a_0}{3}, 
 \end{align}
where $a_S \; (S=0,2)$ is the {\it s}-wave scattering length for the spin channel $S$. In order for the system to be stable, the interaction coefficients should satisfy $c_0 > 0$.

The mean-field energy $E$ is invariant under the $U(1)$ gauge transformation and $SO(3)$ spin rotation, namely, the full symmetry $G$ of the system is given by 
\begin{align}
 G= U(1)_{\phi} \times SO(3)_{\bm{f}}, \label{eq:G}
\end{align}   
where $\phi$ and $\bm{f}$ stand for the gauge and the spin rotation symmetry, respectively. An element $g \in G$ acts on the order parameter $\bm{\psi}$, where $g$ is represented by
\begin{align}
 g  = e^{i \phi} e^{- i f_z \alpha} e^{-i f_y \beta} e^{-i f_z \gamma}.
\end{align}
Here, $\phi$ describes the gauge degree of freedom, and $\alpha$, $\beta$, and $\gamma$ denote the Euler angles in the spin space. 

The ground state of the total energy (\ref{eq:enetot}) is known to be a ferromagnetic (FM) state for $c_1 < 0$ and a polar state for $c_1 >0$, where their representative order parameters are given by
\begin{align}
&\bm{\psi}_{\rm F} = \sqrt{n} (1,0,0)^{T} \ \ (c_1 <0), \label{eq:FMstate}\\
&\bm{\psi}_{\rm P} = \sqrt{\frac{n}{2}}(1,0,1) \ \ (c_1 > 0). \label{eq:Pstate}
\end{align}
These order parameters are invariant under the following operations:
\begin{align}
 H_{\rm F}  &=  \{(e^{i \gamma}, e^{-i f_z \gamma})| \gamma \in [0,2 \pi )\}, \notag \\ 
                  & \cong U(1)_{\phi + f_z}, \label{eq:Hf} \\
 H_{\rm P} &=\{ (1,e^{-i f_y \beta}), (e^{i \pi}, e^{- i f_z \pi} e^{-i f_y \beta} ) | \beta \in [ 0, 2\pi) \}, \notag \\
              & \cong  SO(2)_{f_y} \rtimes (\mathbb{Z}_2)_{\phi + f_z}, \label{eq:Hp}
\end{align}
Here, $H_{\rm F}$ represents the $U(1)$ spin-gauge symmetry and $H_{\rm P}$ is constructed from a rotation symmetry around the $y$ axis in the spin space and the spin-gauge coupled $\mathbb{Z}_2$ symmetry. The subscript $\phi \pm f_z$ means the spin-gauge symmetry in terms of an operation $(e^{i \gamma},e^{\mp f_z \gamma})$. 
The element of $H_{\rm F(P)}$ describes a group action on $\psi $ such that
\begin{align}
 (e^{i \theta}, e^{-i f_{\nu} \gamma}) \bm{\psi} := e^{i \theta} e^{-i f_{\nu} \gamma} \bm{\psi} \ \  ^{\forall} \theta, \gamma \in [0,2 \pi ),
\end{align}
where $\nu = x,y,z$.
Therefore, the OPMs are given by~\cite{Ho:1998,Fei:2001}
 \begin{subequations}
\begin{align}
G/H_{\rm F} &\simeq (U(1)_{\phi} \times SO(3)_{\bm{f}})/U(1)_{\phi + f_z} \notag  \\ &\simeq SO(3)_{\phi - f_z} , \label{eq:Mf}\\
G/H_{\rm P} &\simeq (U(1)_{\phi} \times SO(3)_{\bm{f}})/(SO(2)_{f_y} \rtimes (\mathbb{Z}_2)_{\phi + f_z}) \notag \\ &\simeq (U(1)_{\phi} \times S^2_{\bm{f}})/(\mathbb{Z}_2)_{\phi + f_z}. \label{eq:Mp}
\end{align}
\end{subequations}  

For real systems such as $^{87}$ Rb and $^{23}$Na, the interaction coefficients satisfy $c_0 \gg |c_1|$, which leads to 
\begin{align}
 \frac{c_0 n^2}{2} \gg \frac{|c_1| n^2}{2} | \bm{F}|^2.  \label{eq:c0c1}
\end{align}
Thus, $\epsilon_1$ and $\epsilon_2$ are defined by
\begin{align}
 \epsilon_1 = \frac{c_1 n^2}{2} |\bm{F}|^2, \ \  \epsilon_2 = \frac{c_0 n^2}{2}. 
\end{align} 
The enlarged OPM $\mathcal{M}_2$ is given by minimizing $\epsilon_2$, where the minimum is attained when the local density is constant. In this case, the enlarged OPM is given by
\begin{align}
 \mathcal{M}_2 \simeq S^{5}.
\end{align}
To compare energy scales, we define characteristic lengths for $\epsilon_{\text{kin}} \sim \epsilon_1, \epsilon_2$ as
\begin{align}
 &\xi_{0} = \frac{\hbar}{\sqrt{2 c_0 n M}}, \label{eq:lencond} \\
 &\xi_{\text{s}} = \frac{\hbar}{\sqrt{2 |c_1| n M}}, \label{eq:lenspin}
\end{align}
where $\xi_{0}$ and $\xi_{\text{s}}$ represent the healing length and the spin healing length, respectively. For $c_0 \gg |c_1|$, we have 
\begin{align}
 \xi_{0} \ll \xi_{\text{s}}.
\end{align}
The $r$ dependence of the OPM for $c_0 \gg |c_1|$ is reported in Table~\ref{tab:core}, where $r$ is the distance from the core.

 Since the fundamental group of the enlarged OPM is trivial, i.e., $\pi_1 (S^5) \cong 0$, elements of $\pi_1 (\mathcal{M}_1)$ are always mapped to an identity element:
\begin{align}
 \Phi : \gamma \to 1_{c} ,\ \ ^{\forall} \gamma \in \pi_1 (\mathcal{M}_1), 
\end{align}  
that is, vortices are nonsingular. In fact, by the numerical calculation, the energetically stable vortex-core structures in spin-1 BECs have been achieved by mixing of the FM state and the polar state \cite{Ruostekoski:2003}.   


\begin{table}[tbp]
\centering
 \caption{ The enlarged order parameter manifold in spin-1 BECs in the regions $r \lesssim  \xi_{0}$, $ \xi_0 \lesssim r \lesssim \xi_{\text{s}}$, and  $ \xi_{\text{s}} \lesssim r$. Here, $r$ is the distance from the vortex core, and $\xi_{0}$ and $\xi_{\text{s}}$ are defined by Eqs.~(\ref{eq:lencond}) and~(\ref{eq:lenspin}), respectively. In the rightmost four columns, we list the homotopy group for each enlarged OPM. Homotopy groups $\pi_1$, $\pi_2$, and $\pi_3$ characterize vortices, monopoles (or two-dimensional Skyrmions), and three-dimensional Skyrmions in the three-dimensional space, respectively. The enlarged OPM for the superfluid helium-$3$ is discussed in Ref~\cite{Mineev;1998}. \vspace*{2mm}} \label{tab:core}
 \begin{tabular}{ccccccc} \hline \hline
 Phase & $r$  & (Enlarged) OPM & $\pi_0$ &$\pi_1$ & $\pi_2$ & $\pi_3$ \\ \hline
          & $r \lesssim  \xi_{0}$ & $\mathbb{C}^3$ & 0 & 0 & 0&0 \\
          & $ \xi_{0} \lesssim r \lesssim \xi_{\text{s}}$ & $S^5$ & 0 & 0 & 0 & 0 \\
   FM  & $ \xi_{\text{s}} \lesssim r$ & $SO(3)_{\phi-f_z} $ \cite{Ho:1998} & 0 & $\mathbb{Z}_2$ &  0  & $\mathbb{Z}$\\
    Polar & $ \xi_{\text{s}} \lesssim r$ & $[U(1)_{\phi} \times S^2_{\bm{f}}]/(\mathbb{Z}_2)_{\phi + f_z}$ \cite{Fei:2001} & 0 & $\mathbb{Z}$ & $\mathbb{Z}$ & $\mathbb{Z}$  \\  \hline \hline
 \end{tabular}
\end{table}

\section{General theory for classifying vortex-core structures}
\label{sec:search} 

As we have seen in the preceding section, whether or not a vortex is singular is determined by the inclusion map~(\ref{eq:MVM}). Thus, it is natural to ask what the state inside the vortex core is when the vortex is nonsingular. To investigate the vortex-core structure, we generalize the classification method explained in the previous section.

 We consider an enlarged OPM which does not include a zero point of the order parameter in order to characterize the nonsingular vortex. For such an enlarged OPM, the particle density is nonzero at any point, which implies that the enlarged OPM is homotopic to $S^{2s-1} = \{ \bm{\psi} | \sum_{n=1}^s |\psi_n|^2 = \text{const.}\}$ for a given order parameter $\bm{\psi}= (\psi_1, \cdots , \psi_s) \in \mathbb{C}^s$. Therefore, without loss of generality, we define the maximally enlarged OPM as  
\begin{align}
\mathcal{V} := S^{2s-1}.
\end{align}
The conventional OPM $\mathcal{M}_1$, which is given by minimizing the interaction energy Eq.~(\ref{eq:defOPM}), is included in or equal to the maximally enlarged OPM $\mathcal{V}$:  
\begin{align}
 \mathcal{M}_1 \subseteq \mathcal{V}.
\end{align}
 In the case of spin-$f$ BECs, the maximally enlarged OPM is given by  
\begin{align}
 \mathcal{V} =  S^{4f +1}.
\end{align}

In the following discussion, we classify the core states and the structure of the order parameter inside the vortex core.
We first decompose the maximally enlarged OPM $\mathcal{V}$ according to the symmetry property. Let $G$ be the full symmetry group of the system under consideration, and we consider a $G$ action on $\mathcal{V}$. The {\it $G$-orbit} of $\psi \in \mathcal{V}$ is defined by 
\begin{align}
 [\psi ] = \{ g \psi | g \in G\}.
\end{align}
A set of all orbits in $\mathcal{V}$ is called an {\it orbit space}, which is denoted $\hat{\mathcal{V}}$ in this paper:
\begin{align}
 \hat{\mathcal{V}} := \mathcal{V}/G = \{ [\psi] | \psi \in \mathcal{V} \}. \label{eq:hatV}
\end{align} 
 In addition, we define an {\it isotropy group} of $\psi \in \mathcal{V}$ by Eq.~(\ref{eq:defH}).
 From the definition of the $G$ orbit, elements of the $G$ orbit $[\psi]$ are constructed by applying $g \in G$ to a representative order parameter $\psi$. Since $\psi$ is invariant under the isotropy group $H_{\psi}$,  we should subtract $H_{\psi}$ from $G$ to obtain the $G$ orbit. That is, the $G$ orbit $[\psi]$ is equivalent to $G/H_{\psi}$: 
\begin{align}
 [\psi] \simeq G/H_{\psi} \label{eq:G|H}.
\end{align} 
When $\psi$ is the ground state that minimizes the interaction energy $\epsilon_{\text{int}}$, $G/H_{\psi}$ is identical to the OPM $\mathcal{M}_1$. If $\psi$ and $\psi'$ belong to the same orbit, their isotropy groups are conjugate with each other, i.e.,
\begin{align}
 H_{\psi} = g H_{\psi'} g^{-1},  \ \ ^{\exists}g \in G . \label{eq:equibH}
\end{align} 
Note, however, that even when $\psi$ and $\psi'$ do not belong to the same orbit, their isotropy group may be conjugate to each other, which implies that different orbits may have the same property. Therefore, we categorize the $G$ orbits according to the conjugacy class of their isotropy groups: 
\begin{align}
 \langle \psi \rangle  := \{ [ \psi' ] \in \hat{\mathcal{V}} |  H_{\psi'} = g H_{\psi} g^{-1}, \ \ ^{\exists}g \in G \}. \label{eq:sp-equib}
\end{align}
Such a category is called {\it a stratum}.
\begin{figure*}[tb]
\centering
\includegraphics[width = 14 cm]{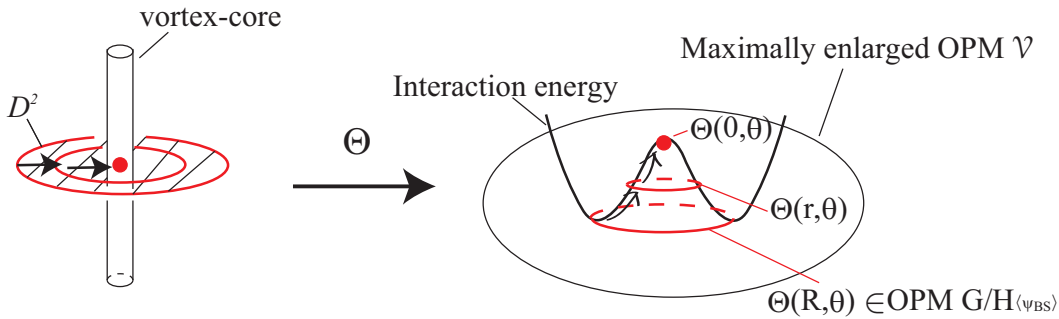}
\caption{(Color online)We consider two-dimensional disk $D^2$ covering a cylindrically symmetric vortex configuration in a real system, where the core is placed at the center and the radius $R$ is much greater than the healing length. Disk $D^2$ is mapped into the maximally enlarged OPM $\mathcal{V}$, where the boundary $\partial D^2$ is mapped onto the conventional OPM denoted $\Theta (R,\theta)$ with $\theta \in [0,2 \pi)$. With decreasing $r$ from $R$ to $0$,  the interaction energy increases because the loop enclosing the vortex gradually moves out of the OPM in order to shrink to a point. Finally, $\Theta (0,\theta)$ is mapped onto the core state in the enlarged OPM.}\label{fig:mapD}
\end{figure*}
Then, the orbit space $\hat{\mathcal{V}}$ is described as a union of $\langle \psi \rangle$:
\begin{align}
 \hat{\mathcal{V}} = \bigcup_{ \psi \in \mathcal{V}} \langle \psi \rangle. \label{eq:hatV2}
\end{align}
 If $\mathcal{V}$ is compact, number of the strata is finite. Moreover, since all isotropy groups for the order parameter in the same stratum are isomorphic to each other, $H_{\psi}$ can be replaced with a representative one $H_{\langle \psi \rangle}$. Then, from Eqs.~ (\ref{eq:hatV}) and (\ref{eq:hatV2}), we obtain
\begin{align}
 \mathcal{V} \simeq \bigcup_{\psi \in \mathcal{V}} (G/H_{\langle \psi \rangle} \times \langle \psi \rangle ). \label{eq:decompV}
\end{align} 

 Next, we consider a map from a two-dimensional disk $D^2$ with radius $R$ to $\mathcal{V}$:
\begin{align}
\Theta: D^2 \to \mathcal{V}. \label{map:D2-1}
\end{align}
Here, we assume that the vortex is placed at the center of $D^2$ and the configuration of the order parameter on $D^2$ is axisymmetric. In this case, $D^2$ is parametrized by $(r, \theta)$ with $ r \in [0,R]$ and $\theta \in [0, 2 \pi)$. The boundary $\partial D^2$ is defined by 
\begin{align}
\partial D^2 = \{ (R, \theta)| \theta \in [0,2 \pi )\}.
\end{align}
The map $\Theta$ is shown schematically in Fig.~\ref{fig:mapD}. Let $\psi_{\text{BS}} \in [\psi_{\text{BS}}]$ be a representative order parameter that minimizes the interaction energy $\epsilon_{\rm int}$, where BS stands for the boundary state; i.e., the OPM is given by 
\begin{align}
 \mathcal{M}_1 \simeq G/H_{\langle \psi_{\text{BS}}\rangle}.
\end{align}

When $R$ is sufficiently large compared with the healing length $\hbar/\sqrt{2M|\epsilon_1|}$, the image of the boundary $\partial D^2$ should be a loop in the OPM,
\begin{align}
 \Theta (R,\theta) \in \mathcal{M}_1\simeq  G/H_{\langle \psi_{\text{BS}} \rangle }, \ \ ^\forall \theta \in [0,2 \pi). 
\end{align}
The homotopy equivalent class of loop $\Theta(R,\theta)$, which is an element of the fundamental group $\pi_1(G/H_{\langle \psi_{\text{BS}} \rangle })$, characterizes the vortex structure away from the core.

To see the vortex-core structure, we need to investigate how loop $\Theta(r,\theta)$ moves in $\mathcal{V}$ as $r$ goes to $0$.
We note that even when $\Theta(R,\theta)$ is fixed in $\mathcal{M}_1$, there are several possibilities for the map $\Theta(r,\theta)\in \mathcal{V} \; (0<r<R)$, which is continuously deformed from $\Theta(R,\theta)$.
 From the assumption that the configuration of the order parameter is axisymmetric about the vortex line, loop $\Theta (r,\theta)$ for a fixed $r$ belongs to a $G$ orbit $[\psi]$:
\begin{align}
\Theta (r,\theta )  \in [\psi] \simeq G/H_{ \langle \psi \rangle },  \ \ ^{\forall} \theta \in [0,2 \pi), \label{eq:asm}  
 \end{align}
 Therefore, the loop $\Theta(r,\theta)$ for every $r$ is classified with the fundamental group $\pi_1 (G/H_{\langle \psi \rangle})$.

 Since $G/H_{\langle \psi \rangle }$ is invariant as long as $\psi$ belongs to the same stratum, $\Theta (r_1,\theta)$ is homotopic to $\Theta (r_2 ,\theta)$ if $[\psi(r)]$ does not change the stratum for $r_1 < r <r_2$. In the case, the homotopy equivalence class of $\Theta(r_1 , \theta )$ and that of $\Theta(r_2 , \theta)$ belong to the same element of $\pi_1 (G/H_{\langle \psi \rangle })$. Accordingly, when $r$ varies from $R$ to $0$, a nontrivial change of $\Theta(r, \theta)$ may occur when it crosses the boundary of strata. The nontrivial deformation of a loop can be characterized by a map between the fundamental groups $\pi_1(G/H_{\langle \psi \rangle})$ of the different strata. Hence, the vortex-core structure is characterized by a sequence of strata and by the map between their fundamental groups.  

In the following discussion, we define a map between the fundamental groups at the boundary of strata. We start by introducing the universal covering space of $G$, which we denote as $\tilde{G}$.
 The universal covering space is a simply connected space, namely, it satisfies
 \begin{subequations}
\begin{align}
 &G \subseteq \tilde{G} , \\
 &\pi_1 (\tilde{G}) \cong 0.
\end{align}
\end{subequations}
If $G$ is a simply connected space, we define $G = \tilde{G}$. Associated with the lift from $G$ to $\tilde{G}$, the isotropy group $H_{\langle \psi \rangle}$ is also lifted to $\tilde{H}_{ \langle \psi \rangle}$ so as to satisfy 
\begin{align}
 G/H_{\langle \psi \rangle} \simeq \tilde{G}/\tilde{H}_{\langle \psi \rangle}.
\end{align} 
 Note that $\tilde{H}_{\langle \psi \rangle}$ is not simply connected. 
When $\tilde{G}$ is simply connected, the fundamental group satisfies the relations 
\begin{align}
\pi_1 (\tilde{G}/\tilde{H}_{\langle \psi \rangle }) \cong \pi_0(\tilde{H}_{\langle \psi \rangle }) \cong \tilde{H}_{\langle \psi \rangle }/(\tilde{H}_{ \langle \psi \rangle })_0, \label{eq:0-1}
\end{align}
where $(\tilde{H}_{\langle \psi \rangle })_0$ is a set of components that are connected to the identity element. Therefore, the map between $\pi_1$'s is defined as a map between $\tilde{H}_{\langle \psi \rangle }$'s.  We further note that the isotropy group of one of adjoining strata is bigger than or smaller than the isotropy group of the other, as proved by Michel~\cite{Michel;1980}.  
It then follows that, when we consider a map from $H_{\langle \psi \rangle}$ to $H_{\langle \psi' \rangle}$, there are only two possibilities: 
(a) $H_{\langle \psi \rangle } \subsetneq  H_{\langle \psi' \rangle}$, and (b) $H_{\langle \psi \rangle } \supsetneq H_{\langle  \psi' \rangle} $. If $H_{\langle \psi \rangle } \subsetneq H_{\langle \psi' \rangle } $ ($H_{\langle \psi \rangle } \supsetneq H_{\langle \psi' \rangle } $), the lifted isotropy group also satisfies $\tilde{H}_{\langle \psi \rangle } \subsetneq \tilde{H}_{\langle \psi' \rangle } $ ($\tilde{H}_{\langle \psi \rangle } \supsetneq \tilde{H}_{\langle \psi' \rangle } $). 
Then, the map from $\tilde{H}_{\langle \psi \rangle }$ to $\tilde{H}_{\langle \psi' \rangle }$ is defined as follows.
\begin{itemize}
 \item[(a)] For the case of $\tilde{H}_{\langle \psi \rangle } \subsetneq  \tilde{H}_{\langle \psi' \rangle} $, the map from $\tilde{H}_{\langle \psi \rangle }$ to $\tilde{H}_{\langle \psi' \rangle }$ is defined by an inclusion map,
 \begin{subequations}
\begin{align}
 i : &  \tilde{H}_{\langle \psi \rangle }\to \tilde{H}_{\langle \psi' \rangle} , \label{map:i-1}\\
     &  a \mapsto a , \ \  ^{\forall} a \in \tilde{H}_{\langle \psi \rangle }. \label{map:i-2}   
\end{align} 
\end{subequations}

 \item[(b)] For the case of $\tilde{H}_{\langle \psi \rangle } \supsetneq \tilde{H}_{\langle \psi' \rangle} $, 
we first decompose $\tilde{H}_{\langle \psi \rangle}$ into a set of connected spaces $\{ X_i\}$ such that
\begin{align}
 \tilde{H}_{\langle \psi \rangle } = \bigcup_{i \in J} X_i,  \ \  X_i \cap X_j = \emptyset, \ \ ^{\forall}i,j \in J,
\end{align}
where $J$ is a set of indices labeling $X_i$. When $H_{\langle \psi \rangle}$ is a connected space, $J$ includes only one element. We then define a map from each $X_i$ to $H_{\langle \psi \rangle}$ as a constant map to an element $b$ in the intersection of $X_i$ and $H_{\langle \psi' \rangle}$:
\begin{subequations} 
 \begin{align}
 c_b^{(i)} :  &X_i \to \tilde{H}_{\langle \psi' \rangle} , \ \ b \in X_i \cap \tilde{H}_{\langle \psi' \rangle}, \label{map:chi-1}\\
       &  a \mapsto  b, \ \ \ \ ^{\forall}a \in X_i. \label{map:chi-2}
\end{align} 
\end{subequations}
For each $X_i$, the number of constant maps is equivalent to that of elements of $X_i \cap \tilde{H}_{\langle \psi' \rangle }$~\cite{text2}.
\end{itemize} 
The map between $\pi_0(\tilde{H}_{\psi})$ and $\pi_0 (\tilde{H}_{\psi'})$ is induced by maps (\ref{map:i-1}) and (\ref{map:chi-1}):
\begin{subequations}
\begin{align}
& i_{\ast} ([a]) := [i(a)], \ \ ^{\forall}a \in \tilde{H}_{\langle \psi \rangle}, \label{map:indi}\\
&(c_b^{(i)})_{\ast} ([a]) := [c_b^{(i)}(a)], \ \ ^{\forall}a \in X_i. \label{map:indch}
\end{align}
\end{subequations}
where $[a]$ denotes an element of $\tilde{H}_{\langle \psi \rangle}/(\tilde{H}_{\langle\psi \rangle})_0$.
Note that map $i$ is determined uniquely and its induced map is a homomorphism, whereas map $c_b^{(i)}$ is not a homomorphism. 

The vortex-core structure is specified by giving all states between $r=0$ and $r=R$ and assigning the winding number to each state. Mathematically, this implies that the vortex-core structure is defined by the order of strata between $r=0$ and $r=R$ and by the map between the lifted isotropy groups at each boundary between adjacent strata given in Eqs.~(\ref{map:indi}) and~(\ref{map:indch}).    
Suppose that as $r$ decreases, the order parameter moves on the strata along the following sequence: 
\begin{align}
 \{ \langle \psi_{\text{BS}} \rangle \to \langle \psi_{1} \rangle \to \langle \psi_2 \rangle \to \cdots \to \langle \psi_{l-1} \rangle \to \langle \psi_{l} \rangle \}. \label{eq:order} 
\end{align}
Accordingly, the OPM changes from $ G/H_{\langle \psi_{\text{BS}} \rangle} $ to $G/H_{\langle \psi_1 \rangle }$, and from $G/H_{\langle \psi_1 \rangle } $ to $G/H_{\langle \psi_2 \rangle}$, and so on. Then, the vortex-core structure is defined as a sequence of maps between the fundamental groups of $G/H_{\langle \psi \rangle }$, which is equivalent to the sequence
 \begin{align}
 \pi_0 (\tilde{H}_{\text{BS}}) \xrightarrow{\Omega_{1}}  \pi_0 (\tilde{H}_1) & \xrightarrow{\Omega_2} \cdots \xrightarrow{\Omega_{l-1}} \pi_0 (\tilde{H}_{l-1}) \notag \\  &\xrightarrow{\Omega_{l}} \pi_0 (\tilde{H}_{l}), \label{eq:sequ}
 \end{align}
where $\tilde{H}_k := \tilde{H}_{\langle \psi_k \rangle} \; (1 \le k \le l)$ is the lifted isotropy group of $H_{\langle \psi_k \rangle }$ and $\Omega_k$ is a map defined in Eq.~(\ref{map:indi}) or~(\ref{map:indch}). In order for loop $\Theta (r , \theta)$ to shrink in the manifold $G/H_{l}$ at $r=0$, sequence~(\ref{eq:sequ}) should satisfy 
\begin{subequations}
\begin{align}
&\Omega_{l-1} \circ \Omega_{l-2} \circ \cdots \circ \Omega_1 ( [h] ) \neq [e], \label{eq:omega1} \\
&\Omega_{l} \circ \Omega_{l-1} \circ \cdots \circ \Omega_1 ( [h] ) = [e] , \label{eq:omega2}
\end{align}
\end{subequations}
 where $[h]$ is the element of $\pi_0 (\tilde{H}_{\text{BS}})$ that characterizes the vortex under consideration, $e$ is the identity element of $\tilde{H}_{l}$ and the composition between maps is defined by
\begin{align} 
\Omega_k \circ \Omega_{k-1} ([h]) = \Omega_k ( \Omega_{k-1} ([h]) ). \label{mapcomp}
\end{align}
 We call the final state of sequence, i.e. $\langle \psi_l \rangle$, a { \it core state} and the sequence in Eq.~(\ref{eq:sequ}) a {\it core structure}. In the following sections, we apply the above classification method to a spin-1 BEC.

\section{Vortex-core structure in a spin-1 BEC}
\label{sec:vortex}
\subsection{Decomposition of $S^5$}
\label{sec:fib}

From Sec.~\ref{sec:applspin1}, the maximally enlarged OPM is $S^5$ in a spin-1 BEC. In this section, we introduce the $G$ orbit and the orbit space of $S^5$.

 The order parameter in the spin-1 BEC is given by Eq.~(\ref{eq:opspin1}), which includes six independent variables. The full symmetry is given by Eq. (\ref{eq:G}); i.e., the energy functional is invariant under a gauge transformation and a spin rotation in the spin space. 
  
To derive the orbit space of $S^5$, we consider a transformation of variables. The physical quantities that characterize the state of the spin-1 BEC are the number density $n$ and the magnetization $|\bm{F}|$. It can be shown that these quantities, overall phase $\phi$, and Euler angles $\alpha$, $\beta$, and $\gamma$ completely specify the spin-1 order parameter. In general, the state of the spin-1 BEC is described by
\begin{align}
 \bm{\psi} = \sqrt{n} e^{i \phi} e^{- i f_z \alpha} e^{ - i f_y \beta} e^{- i f_z \gamma} \left( \begin{array}{@{\,} c @{\,}} \sqrt{\frac{1 + | \bm{F}|}{2}} \\ 0 \\ \sqrt{\frac{1 - | \bm{F}|}{2}}\end{array} \right) . \label{thm:op}
\end{align}  
where $\theta , \alpha , \gamma \in [0, 2 \pi)$, $\beta \in [0, \pi ]$, and $ 0 \le | \bm{F}| \le 1$. The derivation of Eq.~(\ref{thm:op}) is shown in Appendix~\ref{sec:appA}. 
Equation~(\ref{thm:op}) tells us that the orbit space is spanned by $n$ and $|\bm{F}|$.
\begin{figure}[tb]
\centering
\includegraphics[width = 9 cm]{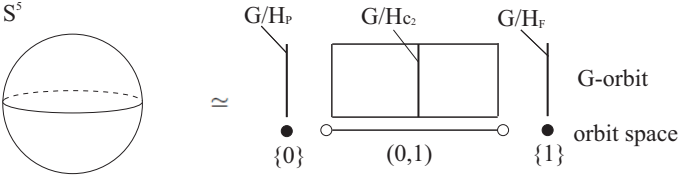}
\caption{Decomposition of $S^5$. By Eq.~(\ref{eq:decompS5}), $S^5$ is homeomorphic to $(G/H_{\rm F} \times \{ 1\}) \cup (G/H_{C_2} \times (0,1)) \cup (G/H_{\rm P} \times \{ 0\})$. The orbit space is given by $[0,1]$, where $[0,1]$ represents the amplitude of local magnetization. The orbit space is decomposed into three strata: $\{ 0\}$, $(0,1)$, and $\{ 1\}$. The corresponding $G$ orbits are given by $G/H_{\rm F}$ at $\{ 0\}$, $G/H_{C_2}$ at $(0,1)$, and $G/H_{\rm P}$ at $\{ 1\}$. }\label{fig:FB}
\end{figure}
When $n$ is a constant, the total degrees of freedom of the order parameter is equal to those of the five-dimensional sphere $S^5$, which corresponds to the maximally enlarged OPM. Thus, the orbit space is parametrized by the local magnetization $0 \le |\bm{F}| \le 1$, and, hence, given by 
\begin{align}
 \hat{\mathcal{V}} = [0,1].
\end{align}
The orbit space $\hat{\mathcal{V}}$ is decomposed into three strata:
\begin{subequations}
\begin{align}
 &\langle \psi_{\rm P}\rangle \simeq \{ 0\}, \\
&\langle \psi_{C_2}\rangle \simeq (0,1), \\
&\langle \psi_{\rm F}\rangle \simeq \{ 1\},
\end{align}
 \end{subequations}
 corresponding to the polar state $(|\bm{F}| = 0)$, the $C_2$ state $(0< |\bm{F}| <1)$~\cite{Yuki:2011}, and the FM state $(|\bm{F}| = 1)$. The order parameter and the isotropy group for the FM state are given by Eqs.~(\ref{eq:FMstate}) and~(\ref{eq:Hf}), respectively, and those for the polar state are given by Eqs.~(\ref{eq:Pstate}) and~(\ref{eq:Hp}). On the other hand, the $C_2$ state is defined as a partially polarized state whose representative order parameter is given by 
\begin{align}
 \bm{\psi}_{C_2} =  \sqrt{n} \left( \sqrt{\frac{1 + | \bm{F}|}{2}} , 0 , \sqrt{\frac{1 - | \bm{F}|}{2}} \right)^T \; (0< |\bm{F}| <1). \label{eq:C2state}
\end{align}
The $C_2$ state is invariant under a $\pi$ rotation about the $z$ axis together with a $\pi$ gauge transformation, i.e., the isotropy group of the $C_2$ state is given by
\begin{align}
 H_{C_2} = \{ (1,\bm{1}_{3}), (e^{i \pi} ,e^{- i f_z \pi}) \} \cong (\mathbb{Z}_2)_{\phi + f_z}. \label{eq:Hc2}
\end{align}
Here, $\bm{1}_{n}$ denotes an $n \times n$ unit matrix. Note that the $C_2$ state cannot be the ground state of the system in the absence of a magnetic field (the magnetic field specifies not only the magnitude but also the direction of the spontaneous magnetization~\cite{Murata:2007}). In other words, contrary to the FM and polar states which become the ground states for $c_1 < 0$ and $c_1 > 0$, respectively, the $C_2$ state involves excitations such as vortices.

By the isotropy groups $H_{\rm F}$ and $H_{\rm P}$, OPMs for the FM and the polar states are given in Eqs. (\ref{eq:Mf}) and (\ref{eq:Mp}), respectively. By Eq.~(\ref{eq:Hc2}), the coset space for the $C_2$ state is given by
\begin{align}
G/H_{C_2} &\simeq (U(1)_{\phi} \times SO(3)_{\bm{f}})/(\mathbb{Z}_2)_{\phi + f_z}, \label{eq:Mc2}
\end{align}
for fixed $|\bm{F}|$.

From the above discussion, we find that $S^5$ can be decomposed in the form of Eq.~(\ref{eq:decompV}). When the local density is constant, $S^5$ is decomposed as follows:
\begin{align}
 S^5 \simeq (G/H_{\rm P} \times \{ 0\}) \cup (G/H_{C_2} \times (0,1)) \cup (G/H_{\rm F} \times \{ 1\}), \label{eq:decompS5}
\end{align}
where $[0,1]$ represents the amplitude of local magnetization $|\bm{F}|$. We show the decomposition of $S^5$ graphically in Fig.~\ref{fig:FB}. 

\subsection{Construction of maps}
\label{sec:maps}
We calculate the fundamental group of each state and derive the map $i$ and $c_b^{(i)}$ defined in Eqs.~(\ref{map:i-2}) and~(\ref{map:chi-2}), respectively. We begin by calculating elements of the fundamental group using Eq.~(\ref{eq:0-1}). 
 Since the symmetry group $G= U(1)_{\phi} \times SO(3)_{\bm{f}}$ in the spin-1 BEC is not simply connected, we lift it to its universal covering space:
\begin{align}
 \tilde{G} = \mathbb{R}_{\phi} \times SU(2)_{\bm{f}}. \label{eq:liftG}
\end{align}
Associated with the lift, the isotropy group $H_q \; (q = {\rm F,P},C_2)$ is also lifted to $\tilde{H}_q$.
 As a result, $\tilde{H}_{\rm F}$, $\tilde{H}_{\rm P}$, and $\tilde{H}_{C_2}$ are given by
\begin{subequations}
 \begin{align}
 \tilde{H}_{\rm F} &:= \{ (x , \pm e^{ - i \frac{\sigma_z}{2} x}) | x \in \mathbb{R} \}, \label{tildeHa} \\
 \tilde{H}_{\rm P} &:= \left\{ (\pi n ,e^{-i \frac{\sigma_z}{2} \pi n} e^{ - i \frac{\sigma_y}{2} \beta }) \Big| \;  n \in \mathbb{Z}, \; \beta \in [0,4\pi) \right\}, \label{tildeHb} \\
  \tilde{H}_{C_2} &:= \left\{ (  \pi n, \pm e^{-i \frac{\sigma_z}{2} \pi n} ) \Big| \; n \in \mathbb{Z} \right\}, \label{tildeHc}
 \end{align}
\end{subequations}
where $\sigma_{\nu} , \; (\nu=x,y,z)$ are the Pauli matrices. The derivations of Eqs. (\ref{tildeHa}), (\ref{tildeHb}), and (\ref{tildeHc}) are given in Appendix~\ref{sec:appB}. Here, we add the factor $\pi $ in front of $n \in \mathbb{Z}$ for the sake of convenience.  In the following discussion, we describe each lifted isotropy group by the representation of Eqs. (\ref{tildeHa})--(\ref{tildeHc}).

\begin{figure*}[tb]
\centering
\includegraphics[width = 14 cm]{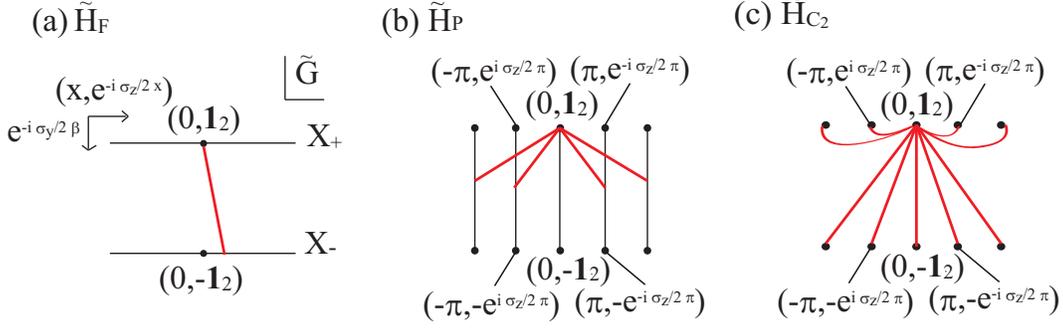}
\caption{(Color online) Structure of isotropy groups (a) $\tilde{H}_{\rm F}$, (b) $\tilde{H}_{\rm P}$, and (b)$\tilde{H}_{C_2}$ in the universal covering group $\tilde{G}$. The vertical axis shows the transformation of $SU(2)_{\bm{f}}$ represented by $e^{- i \frac{\sigma_y}{2 } \beta}$ with $\beta \in [0,4 \pi )$, and the horizontal axis represents the transformation of $\mathbb{R}_{\phi} \times SU(2)_{\bm{f}}$ given by $(x, e^{-i \frac{\sigma_z}{2} x})$ with $x \in \mathbb{R}$. (a) $\tilde{H}_{\rm F}$ is composed of two connected spaces, $X_+ = \{ (x ,e^{- i \frac{\sigma_z}{2} x}) | x \in \mathbb{R}\}$ and $X_- = \{ (x ,-e^{- i \frac{\sigma_z}{2} x}) | x \in \mathbb{R}\}$, which correspond to the upper and lower horizontal lines, respectively. (b) $\tilde{H}_{\rm P}$ is composed of $X_{n_p} = \{( \pi n_p , e^{-i \frac{\sigma_z}{2} \pi n_p}e^{ - i \frac{\sigma_y}{2} \beta }) | \beta \in [0,4 \pi )\}$, where $n_p \in \mathbb{Z}$. Each $X_{n_p}$ corresponds to a vertical line. (c) $\tilde{H}_{C_2}$ consists of discrete elements $X_{(+ , n_p)}=( \pi n_p,  e^{-i \frac{\sigma_z}{2} \pi n_p})$ and $X_{(- , n_p)}=( \pi n_p,  - e^{-i \frac{\sigma_z}{2} \pi n_p})$ with $n_p \in \mathbb{Z}$, which are represented by filled circles. The thick (red) lines represent the nontrivial loop in the coset space $\tilde{G}/\tilde{H}_{q} \; (q = {\rm F, P}, C_2)$. Since all elements of $\tilde{H}_q$ are identified in the coset space, a loop in $G/H_q$ is described by a segment that connects the element of $\tilde{H}_q$. From this figure, we can understand nontrivial loops such as (a) a unique loop, (b) an infinite number of loops labeled by single integers, and (c) an infinite number of loops labeled by two integers.}\label{fig:Hf}
\end{figure*}

In the FM state, a set of components connected to $(0,\bm{1}_2)$, $(\tilde{H}_{\rm F})_0$, is given by 
\begin{align}
 (\tilde{H}_{\rm F})_0 = \{ (x , e^{ - i \frac{\sigma_z}{2} x}) | x \in \mathbb{R} \}. \label{eq:Hf0}
\end{align}
 Hence, the zeroth homotopy group of the FM state is given by
\begin{align}
\pi_0 (\tilde{H}_{\rm F}) \cong \tilde{H}_{\rm F} / (\tilde{H}_{\rm F})_0 &= \{ (0,\bm{1}_{2}) , (0, - \bm{1}_{2})\} \notag \\&\cong \mathbb{Z}_2. \label{vor:2-4}
\end{align}
The nontrivial element $(0, - \bm{1}_{2})$ represents a spin vortex. The structure of $\tilde{H}_{\rm F}$ is  illustrated schematically in Fig. \ref{fig:Hf} (a). 

 For the polar state, $(\tilde{H}_{\rm P})_0 $ is given by
\begin{align}
(\tilde{H}_{\rm P})_0 = \{ (0, e^{-i \frac{\sigma_y}{2} \beta}) | \beta \in [ 0 , 4 \pi )\}, \label{vor:2-5}
\end{align}
from which the zeroth homotopy group is given by
\begin{align}
 \pi_0 (\tilde{H}_{\rm P}) & \cong  \tilde{H}_{\rm P}/ (\tilde{H}_{\rm P})_0 \notag \\ 
                          &= \{ (2 n\pi,  \bm{1}_{2}), ( (2 n + 1) \pi,  e^{-i \frac{\sigma_z}{2} \pi} ) | n \in \mathbb{Z}\}  \notag \\
                          & \cong \mathbb{Z} . \label{vor:2-6}
\end{align}
Hence, vortices in the polar state are classified by an integer: $(2 \pi n,  \bm{1}_{2})$ describes a gauge vortex, whereas $ ( (2 n + 1) \pi,  e^{-i \frac{\sigma_z}{2} \pi} )$ represents a half-integer vortex (or Alice vortex). The structure of $\tilde{H}_{\rm P}$ is shown in Fig \ref{fig:Hf} (b). 

For the $C_2$ state, since $\tilde{H}_{C_2}$ is a discrete subgroup of $\tilde{G}$, $(\tilde{H}_M)_0$ is composed of the identity element alone;
\begin{align}
 (\tilde{H}_{C_2})_0 = \{ (0,\bm{1}_{2}) \}.
\end{align}
 Therefore, the zeroth homotopy group is isomorphic to $\tilde{H}_{C_2}$: 
\begin{align}
 \pi_0 (\tilde{H}_{C_2}) & \cong \tilde{H}_{C_2} \notag \\ 
                      &=  \left\{ (2 \pi n, \pm \bm{1}_{2}),( (2  n + 1) \pi, \pm e^{-i \frac{\sigma_z}{2} \pi}) \Big| n \in \mathbb{Z} \right\}.  \label{vor:2-7}
\end{align}
Comparing Eq.~(\ref{vor:2-7}) with Eqs.~(\ref{vor:2-4}) and~(\ref{vor:2-6}), $\pi_0(\tilde{H}_{C_2})$ includes the elements of $\pi_0(\tilde{H}_{\rm F})$ and those of $\pi_0(\tilde{H}_{\rm P})$. These structures are important to characterize the vortex-core structures as discussed in Sec. \ref{sec:classify}. The structure of $\tilde{H}_{C_2}$ is shown in Fig. \ref{fig:Hf} (c). 
   
 Let us construct the map between $\pi_0$'s of each lifted isotropy group. The lifted isotropy groups $\tilde{H}_{\rm F}$, $\tilde{H}_{\rm P}$, and $\tilde{H}_{C_2}$ satisfy 
 \begin{align}
  \tilde{H}_{C_2} \subset \tilde{H}_{\rm F},\tilde{H}_{\rm P}. \label{eq:relH}
 \end{align}  
  Since $\langle \psi_{\rm F(P)} \rangle$ does not adjoin $\langle \psi_{\rm P(F)} \rangle$, we have to pass through the $C_2$ state. Therefore, we need to define only four types of maps: (a) from $\tilde{H}_{\rm F}$ to $\tilde{H}_{C_2}$, (b) from $\tilde{H}_{C_2}$ to $\tilde{H}_{\rm F}$, (c) from $\tilde{H}_{\rm P}$ to $\tilde{H}_{C_2}$, and (d) from $\tilde{H}_{C_2}$ to $\tilde{H}_{\rm P}$:
\def\reacteqarrow#1#2#3#4{\raisebox{#1}{\parbox{#2}{\centering{#3} \par \vskip-0.25\normalbaselineskip \rightarrowfill \par \vskip-0.75\normalbaselineskip \leftarrowfill \par \vskip-0.25\normalbaselineskip \centering{#4}}}}
\begin{align}
 \tilde{H}_{\rm F} \ \  \reacteqarrow{0cm}{1.2cm}{$c_{n_p}^{(\pm )}$}{$i_{\rm F}$} \ \ \tilde{H}_{C_2} \ \ \reacteqarrow{0cm}{1.2cm}{$i_{\rm P}$}{$c_{n_f}^{(n_p)}$} \ \ \tilde{H}_{\rm P}.
\end{align}
 Each map is denoted $c_{n_f}^{(n_p)}$, $i_{\rm F}$, $c_{n_p}^{(\pm)}$, and $i_{\rm P}$, and defined as follows:
\begin{itemize}

\item[(a)] $c_{n_p}^{(\pm)}$: Since $\tilde{H}_{\rm F} \supset \tilde{H}_{C_2}$ and $\tilde{H}_{\rm F}$ consists of two connected spaces $X_{+} =\{  (x,  e^{-i \frac{\sigma_z}{2} x} ) | x \in \mathbb{R}\}$ and $X_{-} = \{ (x, - e^{-i \frac{\sigma_z}{2} x}) | x \in \mathbb{R} \}$, the map is defined by Eq.~(\ref{map:chi-1}), i.e.,
\begin{align}
 c_{n_p}^{(\pm)} & : \tilde{H}_{\rm F} \to \tilde{H}_{\rm C_2} , \\
          & (x, \pm e^{-i \frac{\sigma_z}{2} x}) \mapsto (  \pi n_p , \pm e^{-i \frac{\sigma_z}{2} \pi n_p}),
\end{align}
where $x \in \mathbb{R}$ and $n_p \in \mathbb{Z}$ specifies an element of $\tilde{H}_{C_2}$.

\item[(b)] $i_{\rm F}$: Since $\tilde{H}_{C_2} \subset \tilde{H}_{\rm F}$, the map is given by the inclusion map
\begin{align}
 i_{\rm F} &: \tilde{H}_{C_2} \to \tilde{H}_{\rm F}, \\
              & ( \pi n_p, \pm e^{-i \frac{\sigma_z}{2}  \pi n_p}) \mapsto ( \pi n_p, \pm e^{-i \frac{\sigma_z}{2}  \pi n_p}), \quad n_p \in \mathbb{Z},
\end{align}
where  $i_{\rm F}$ is determined uniquely.   
\item[(c)] $c_{n_f}^{(n_p)}$: Since $\tilde{H}_{\rm P} \supset \tilde{H}_{C_2}$ and $\tilde{H}_{\rm P}$ consists of infinite connected spaces labeled by an integer $n_p$: $X_{n_p} =\{ (2  \pi n_p, e^{-i \frac{\sigma_y}{2} \beta }e^{-i \frac{\sigma_z}{2} 2 \pi n_p }) | \beta \in [0 ,4 \pi ) \}$, the constant map is given by.
\begin{align}
&c_{n_f}^{(n_p)}  : \tilde{H}_{\rm P} \to \tilde{H}_{\rm C_2} , \\
          & (2  \pi n_p, e^{-i \frac{\sigma_y}{2} \beta }e^{-i \frac{\sigma_z}{2} 2 \pi n_p }) \mapsto (2 \pi n_p, e^{-i \frac{\sigma_y}{2} 2 \pi n_f } e^{-i \frac{\sigma_z}{2} 2 \pi n_p}),
\end{align}
where $\beta \in [ 0, 4 \pi )$ and $n_f \in \{ 0, 1 \}$ specifies an element of $\tilde{H}_{C_2}$.
\item[(d)] $i_{\rm P}$: Since $\tilde{H}_{C_2} \subset \tilde{H}_{\rm P}$, the map is given by the inclusion map
 \begin{align}
  i_{\rm P} &: \tilde{H}_{C_2} \to \tilde{H}_{\rm P}, \\
                &( \pi n_p, \pm e^{-i \frac{\sigma_z}{2}  \pi n_p}) \mapsto ( \pi n_p, \pm e^{-i \frac{\sigma_z}{2} \pi n_p }),
\end{align}
where  $i_{\rm P}$ is determined uniquely.
 \end{itemize}
 \begin{figure*}[ptb]
\centering
\includegraphics[width = 16 cm]{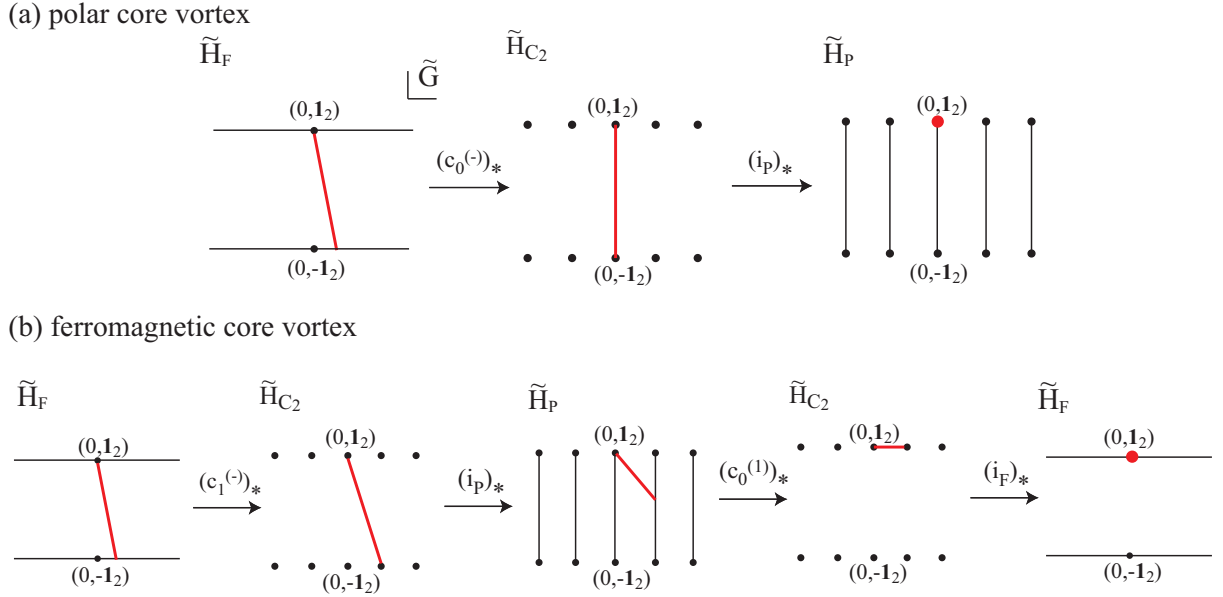}
\caption{(Color online) Deformation processes for (a) the polar-core vortex with $n_p =0 $ and (b) the FM-core vortex with $n_p = 1$ in the FM state. (a) The vortex, which is topologically stable in the FM state, shrinks to a point in the polar state. In the intermediate region, we run through the $C_2$ state. (b) The vortex shrinks to a point in the FM state.  In the intermediate region, we pass through the $C_2$ state and the polar state. The polar state has a nontrivial winding.}\label{fig:classf}
\end{figure*}
 Using these maps $c_{n_p}^{(\pm)}$, $i_{\rm F}$, $c_{n_f}^{(n_p)}$, and $i_{\rm P}$, the maps between $\pi_0$'s are induced as follows:
\begin{subequations}
\begin{align}
 &(c_{n_p}^{(\pm )})_{\ast} : \pi_0 (\tilde{H}_{\rm F}) \to \pi_0(\tilde{H}_{C_2}), \\
&(i_{\rm F})_{\ast} : \pi_0 (\tilde{H}_{C_2}) \to \pi_0(\tilde{H}_{\rm F}), \\
 &(c_{n_f}^{(n_p)})_{\ast} : \pi_0 (\tilde{H}_{\rm P}) \to \pi_0(\tilde{H}_{C_2}), \\
&(i_{\rm P})_{\ast} : \pi_0 (\tilde{H}_{C_2}) \to \pi_0(\tilde{H}_{\rm P}). 
 \end{align}
\end{subequations}
The definition of the induced map is given by Eqs. (\ref{map:indi}) and (\ref{map:indch}).

\subsection{Classification of a vortex-core structure}
\label{sec:classify}
\subsubsection{FM state}
For the case of $c_1 < 0$, the boundary state becomes the FM state,
\begin{align}
 \bm{\psi}_{\text{BS}} = \bm{\psi}_{\rm F} = \sqrt{n} (1,0,0)^T. 
\end{align}
The nontrivial element of $\pi_0 (\tilde{H}_{\rm F})$ is given by $[(0, - \bm{1}_2)]$. Here, $[\cdots ]$ represents the homotopy equivalent class. In what follows, we derive all possible vortex-core structures with the ferromagnetic boundary state. First, by using the map $(c_{0}^{(-)})_{\ast}$, we obtain 
\begin{align}
(i_{\rm P })_{\ast}  \circ (c_{0}^{(-)})_{\ast} ([(0, - \bm{1}_2)]) &= (i_{\rm P })_{\ast}([(0,-\bm{1}_2)])  \notag \\
                                                                               &= [(0,-\bm{1}_2)]  \notag \\ 
                                                                                & = [(0,\bm{1}_2)] \in \pi_0 (\tilde{H}_{\rm P}), \label{eq:fp-1}
\end{align}
where  $(0,\bm{1}_2)$ is the identity element of $\tilde{H}_q$ ($q= {\rm F, P}, C_2$).
For the last equality in Eq.~(\ref{eq:fp-1}), we use the homotopy equivalence relation in $\tilde{H}_{\rm P}$,
\begin{align}
(0, \bm{1}_2) \sim (0, -\bm{1}_2) \in (\tilde{H}_{\rm P})_0. \label{eq:relhp}
\end{align} 
The deformation process in Eq.~(\ref{eq:fp-1}) is illustrated schematically in Fig.~\ref{fig:classf} (a).  Equation~(\ref{eq:fp-1}) shows that there exists a vortex whose core structure is given by
\begin{align}
 \pi_0 (\tilde{H}_{\rm F}) \xrightarrow{(c_{0}^{(-)})_{\ast}}  \pi_0 (\tilde{H}_{C_2})  \xrightarrow{(i_{\rm P})_{\ast}} \pi_0 (\tilde{H}_{\rm P}), \label{thm2-1}
\end{align}
where sequence~(\ref{thm2-1}) satisfies Eqs.~(\ref{eq:omega1}) and~(\ref{eq:omega2}); i.e.,
\begin{subequations}
\begin{align}
 &(c_{0}^{(-)})_{\ast}([(0, -\bm{1}_2)]) = [(0,-\bm{1}_2)], \\
& (i_{\rm P})_{\ast} \circ (c_{0}^{(-)})_{\ast} ([(0, -\bm{1}_2)]) = [(0,\bm{1}_2)]. 
\end{align}
\end{subequations}
According to Eq.~(\ref{eq:order}), we describe the order of strata of Eq. (\ref{thm2-1}) as
\begin{align}
\{\langle \bm{\psi}_{\rm F} \rangle \to \langle \bm{\psi}_{C_2} \rangle \to \langle  \bm{\psi}_{\rm P} \rangle \}. \label{eq:fp-3}
\end{align}
We call Eq.~(\ref{thm2-1}) a polar-core vortex, which was observed experimentally~\cite{Sadler:2006}.
\begin{figure*}[ptb]
\centering
\includegraphics[width = 16 cm]{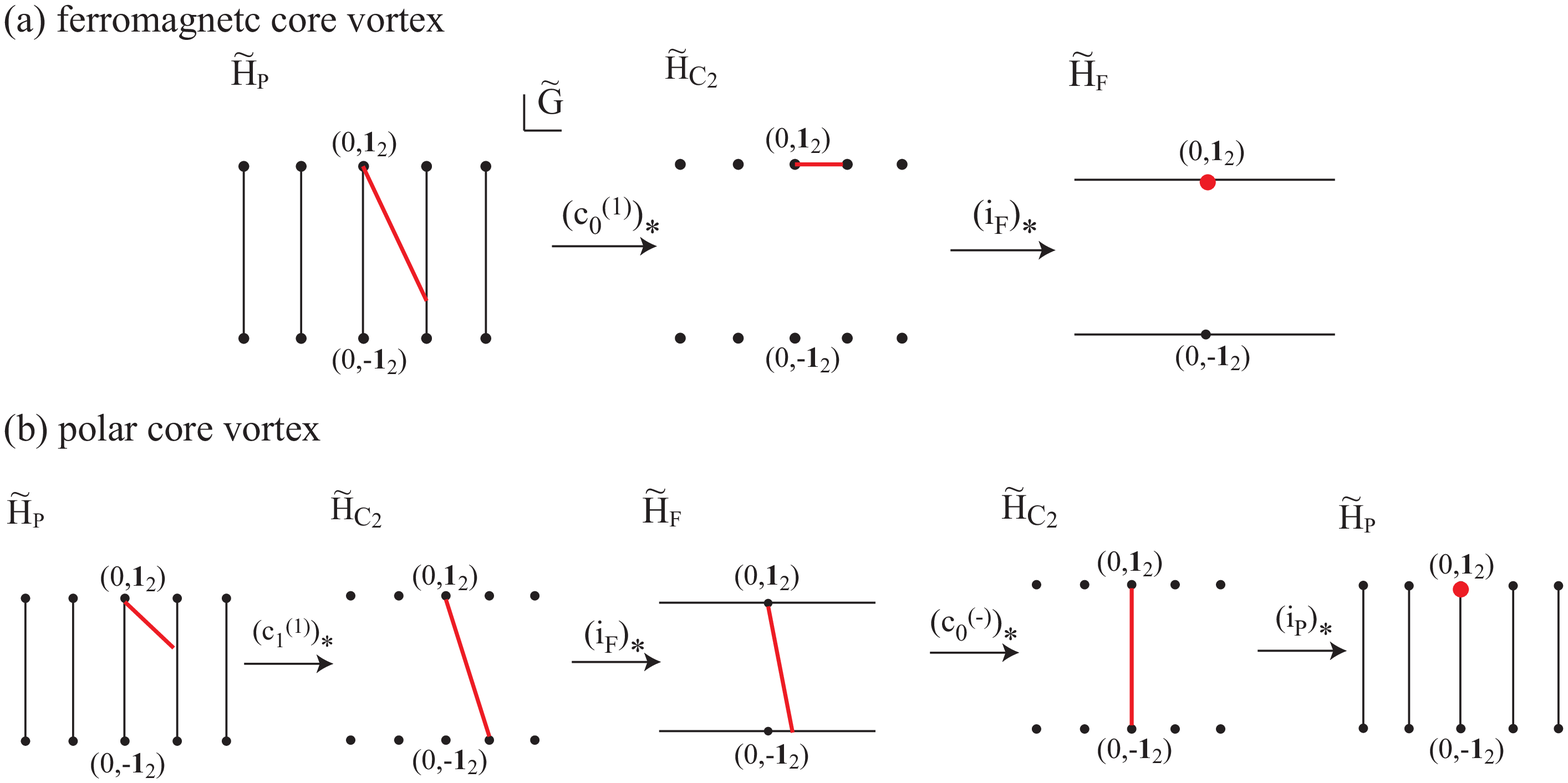}
\caption{(Color online) Deformation process for the case of (a) a FM-core vortex with $n_f =0 $ and (b) a polar-core vortex with $n_f = 1$ in the polar state. (a) The vortex, which is topologically stable in the polar state, shrinks to a point in the FM state. In the intermediate region, we run though the $C_2$ state. (b) The vortex shrinks to a point in the polar state.  In the intermediate region, we pass through the $C_2$ state and the FM state. The FM state has a nontrivial winding.}\label{fig:classp}
\end{figure*}  

Next, by using the map $(c_{n_p}^{(-)})_{\ast} \; (n_p \neq 0)$, we can obtain different types of vortex-core structures. We obtain the following:  
\begin{align}
 (i_{\rm F})_{\ast}  &\circ (c_{0}^{(n_p)})_{\ast} \circ (i_{\rm P })_{\ast}  \circ (c_{n_p}^{(-)})_{\ast} ([(0, - \bm{1}_2)]) \notag \\  
                                                      & = (i_{\rm F})_{\ast}  \circ (c_{0}^{(n_p)})_{\ast} \circ (i_{\rm P })_{\ast} ([( \pi n_p,  -e^{-i \frac{\sigma_z}{2} \pi n_p })]) \notag \\
& = (i_{\rm F})_{\ast}  \circ (c_{0}^{(n_p)})_{\ast} ([( \pi n_p,  -e^{-i \frac{\sigma_z}{2} \pi n_p })])  \notag \\
& = (i_{\rm F})_{\ast}  ([( \pi n_p,  e^{-i \frac{\sigma_z}{2} \pi n_p })])  \notag \\
& = [( \pi n_p,  e^{-i \frac{\sigma_z}{2} \pi n_p })]  \notag \\
& = [(0, \bm{1}_2)] \in \pi_0 (\tilde{H}_{\rm F}), \label{eq:fpf-1}
\end{align}
where the last equality in Eq.~(\ref{eq:fpf-1}) is obtained by the homotopy equivalence relations [see Eq.~(\ref{eq:Hf0})],
\begin{align}
 (0, \bm{1}_2) \sim (n_p \pi ,  e^{-i \frac{\sigma_z}{2} \pi n_p })  \in (\tilde{H}_{\rm F})_0. \label{eq:relhf}
\end{align}
The deformation process of Eq.~(\ref{eq:fpf-1}) is illustrated in Fig.~\ref{fig:classf} (b).
Hence, Eq.~(\ref{eq:fpf-1}) implies that there is a vortex whose core structure is given by 
\begin{align}
 \pi_0 (\tilde{H}_{\rm F}) \xrightarrow{(c_{n_p}^{(-)})_{\ast}}  &\pi_0 (\tilde{H}_{C_2})  \xrightarrow{(i_{\rm P})_{\ast}} \pi_0 (\tilde{H}_{\rm P}) \notag \\ & \xrightarrow{(c_{0}^{(n_p)})_{\ast}}  \pi_0 (\tilde{H}_{C_2})  \xrightarrow{(i_{\rm F})_{\ast}} \pi_0 (\tilde{H}_{\rm F}), \label{thm2-2}
\end{align}
where this sequence~(\ref{thm2-2}) satisfies Eqs.~(\ref{eq:omega1}) and~(\ref{eq:omega2}), i.e.,
\begin{subequations} 
\begin{align}
&(c_{0}^{(n_p)})_{\ast} \circ (i_{\rm P})_{\ast} \circ (p_{n_p}^{(-)})_{\ast} ([0, -\bm{1}_2]) = [( \pi n_p,  e^{-i \frac{\sigma_z}{2} \pi n_p})], \\
&(i_{\rm F})_{\ast} \circ (c_{0}^{(n_p)})_{\ast} \circ (i_{\rm P})_{\ast} \circ (c_{n_p}^{(-)})_{\ast}([(0, -\bm{1}_2)]) =  [(0,\bm{1}_2)].
\end{align}
\end{subequations}
We describe the order of strata as
\begin{align}
\{\langle \bm{\psi}_{\rm F}  \rangle \to \langle \bm{\psi}_{C_2} \rangle \to \langle \bm{\psi}_{\rm P} \rangle \to \langle \bm{\psi}_{C_2} \rangle \to \langle \bm{\psi}_{\rm F} \rangle \}. \label{eq:fpf-3}
\end{align}
Since the core state is the FM state, we call Eq.~(\ref{thm2-2}) a FM-core vortex, which is considered as a new vortex-core structure.
From Eqs.~(\ref{thm2-1}) and (\ref{thm2-2}), we can achieve the vortex-core state when the map $(c_0^{(-)})_{\ast}$ or $(c_0^{(n_p)})_{\ast}$ appears in the sequence. Thus, if these maps do not appear, we can construct a long sequence of vortex-core structure such as 
\begin{align}
\{ \langle  \bm{\psi}_{\rm F} \rangle \to \langle \bm{\psi}_{C_2} \rangle \to  \langle \bm{\psi}_{\rm P} \rangle \to \langle \bm{\psi}_{C_2} \rangle  \to  \langle \bm{\psi}_{\rm F} \rangle \to \cdots  \}. \label{eq:long}
\end{align}
However, we do not discuss  such structures further in this paper because the extension to this case is straightforward.   

 As a result, the core structure depends on the first choice of $(c_{n_p}^{(-)})_{\ast}$, and we can interpret $n_p$ as a quantum number of the core. In practice, $n_p$ is equivalent to the winding number along a ring of the polar state in the core. The FM-core vortex is accompanied by a surrounding polar state with a nontrivial winding number $n_p \neq 0$ in the core. We call such a vortex, whose core structure has a nontrivial winding number, an {\it excited core state}. In Sec.~\ref{sec:dyn}, we show that a FM-core vortex is energetically stable under the quadratic Zeeman effect.

\subsubsection{Polar state} 
For the case of $c_1 > 0$, the boundary state becomes the polar state,
\begin{align}
 \bm{\psi}_{\text{BS}} = \bm{\psi}_{\rm P} = \sqrt{\frac{n}{2}} (1,0,1)^T.
\end{align} 
Since $\pi_0 (\tilde{H}_{\rm P}) \cong \mathbb{Z}$, it is possible to create infinite types of vortices. The vortex is characterized by $[( \pi n_p, e^{-i \frac{\sigma_z}{2} \pi n_p})] \in \pi_0 (\tilde{H}_{\rm P})$ $(n_p \in \mathbb{Z})$. In what  follows, we consider a vortex with the winding number $n_p$. First, by using $(c_{0}^{(n_p)})_{\ast}$, we obtain 
\begin{align}
 (i_{\rm F})_{\ast} \circ (c_{0}^{(n_p)})_{\ast} & ([( \pi n_p, e^{-i \frac{\sigma_z}{2}  \pi n_p)}]) \notag \\
 &= (i_{\rm F})_{\ast} ([( \pi n_p, e^{-i \frac{\sigma_z}{2} \pi n_p})]) \notag \\
  &= [( \pi n_p, e^{-i \frac{\sigma_z}{2} \pi n_p})] \notag \\
   & =  [(0, \bm{1}_2)]. \label{eq:pf-2}
\end{align}
The last equality in Eq.~(\ref{eq:pf-2}) is given by the homotopy equivalence relation given in Eq.~(\ref{eq:relhf}). The deformation process Eq.~(\ref{eq:pf-2}) is shown in Fig.~\ref{fig:classp} (a). Therefore, Eq.~(\ref{eq:pf-2}) means that there is a vortex whose core structure is given by
\begin{align}
 \pi_0 (\tilde{H}_{\rm P} ) \xrightarrow{(c_{0}^{(n_p)})_{\ast}} \pi_0 (\tilde{H}_{C_2}) \xrightarrow{(i_{\rm F})_{\ast}} \pi_0 (\tilde{H}_{\rm F}), \label{thm3-1}
\end{align}
where the sequence satisfies Eqs.~(\ref{eq:omega1}) and~(\ref{eq:omega2}):
\begin{subequations}
\begin{align}
 &(c_{0}^{(n_p)})_{\ast}([ (\pi n_p, e^{-i \frac{\sigma_z}{2} \pi n_p})]) = [( \pi n_p, e^{-i \frac{\sigma_z}{2} \pi n_p})], \\
&(i_{\rm F})_{\ast} \circ (c_{0}^{(n_p)})_{\ast} ([( \pi n_p, e^{-i \frac{\sigma_z}{2} \pi n_p})]) = [(0,\bm{1}_2)], \ \ ^{\forall}n_p \in \mathbb{Z}.
\end{align}
\end{subequations}
The order of strata is given by 
\begin{align} 
\{\langle \bm{\psi}_{\rm P} \rangle \to \langle \bm{\psi}_{C_2} \rangle \to \langle  \bm{\psi}_{\rm F} \rangle \}. \label{eq:pf-3}
\end{align}
Accordingly, this type of vortex-core structure has an intermediate state involving the $C_2$ state.
By using $(c_{1}^{(n_p)})_{\ast}$, we obtain
\begin{align}
 (i_{\rm P})_{\ast} &\circ (c_{0}^{(-)})_{\ast} \circ (i_{\rm F})_{\ast} \circ (c_{1}^{(n_p)})_{\ast} ([( \pi n_p, e^{-i \frac{\sigma_z}{2} \pi n_p})]) \notag \\
 &= (i_{\rm P})_{\ast} \circ (c_{0}^{(-)})_{\ast} \circ (i_{\rm F})_{\ast} ([(\pi n_p, -e^{-i \frac{\sigma_z}{2} \pi n_p})]) \notag \\
 &= (i_{\rm P})_{\ast} \circ (c_{0}^{(-)})_{\ast}  ([(\pi n_p, -e^{-i \frac{\sigma_z}{2}\pi  n_p})]) \notag \\
 &= (i_{\rm P})_{\ast} ([(0, - \bm{1})]) \notag \\
&=  [(0, - \bm{1})] \notag \\
 &=  [(0, \bm{1})], \label{eq:pfp-2}
\end{align}
where the final equality is given by the homotopy equivalence relation in Eq.~(\ref{eq:relhp}). The deformation process in Eq.~(\ref{eq:pfp-2}) is illustrated in Fig.~\ref{fig:classp} (b). Hence, there exists a vortex-core structure given by
\begin{align}
 \pi_0 (\tilde{H}_{\rm P} ) \xrightarrow{(c_{1}^{(n_p)})_{\ast}} &\pi_0 (\tilde{H}_{C_2}) \xrightarrow{(i_{\rm F})_{\ast}} \pi_0 (\tilde{H}_{\rm F}) \notag \\ &\xrightarrow{(c_{0}^{(-)})_{\ast}} \pi_0 (\tilde{H}_{C_2}) \xrightarrow{(i_{\rm P})_{\ast}} \pi_0 (\tilde{H}_{\rm P}), \label{thm3-2}
\end{align}
where sequence~(\ref{thm3-2}) satisfies Eqs.~(\ref{eq:omega1}) and~(\ref{eq:omega2}):
\begin{subequations}
\begin{align}
(c_{0}^{(-)})_{\ast} \circ (i_{\rm F})_{\ast} \circ (c_{1}^{(n_p)} )_{\ast}([( \pi n_p, e^{-i \frac{\sigma_z}{2} \pi n_p})]) = [&(0,- \bm{1}_2)], \\
(i_{\rm P})_{\ast} \circ (c_{0}^{(-)})_{\ast} \circ (i_{\rm F})_{\ast} \circ (c_{1}^{(n_p)})_{\ast} ([( \pi n_p, e^{-i \frac{\sigma_z}{2} \pi n_p}&)])  \notag  \\
 = [&(0,\bm{1}_2)].
\end{align}
\end{subequations}
The order of strata is given by
\begin{align}
 \{ \langle \bm{\psi}_{\rm P} \rangle \to \langle \bm{\psi_{C_2}} \rangle \to \langle \bm{\psi}_{\rm F} \rangle \to \langle \bm{\psi}_{C_2} \rangle \to \langle \bm{\psi}_{\rm P} \rangle \}. \label{eq:pfp-3}
\end{align}

Similarly to the FM state, we can prove the extension of Eq.~(\ref{thm3-2}) to the long sequence vortex-core structure, but we only state the result without discussing such a core structure: 
there are two types of vortex-core structures, which depend on which of $ (c_{0}^{(n_p)} )_{\ast}$ and $ (c_{1}^{(n_p)} )_{\ast}$ operates first. Similarly to the FM state, the label $n_f = \{ 0, 1\}$ is interpreted as a quantum number characterizing the core, and it is equivalent to the winding number along a ring of the FM state in the core. Therefore, there exists an excited state of the core in the polar state.

\section{Energetic stability of vortex core states}
\label{sec:dyn}
In this section, we discuss the energetic stability of a vortex-core state. We show that a vortex whose core structure has a nontrivial winding number can be realized experimentally. We consider a system with a negative quadratic Zeeman energy. The mean-field energy with the negative quadratic Zeeman term is given by
\begin{align}
 E = \int d \bm{r} \left( \epsilon_{\text{kin}} + \epsilon_{\text{int}} + \epsilon_{\text{zeeman}} \right), \label{eq:eneztot}
\end{align}
where $\epsilon_{\text{kin}}$ and $\epsilon_{\text{int}}$ are given by Eqs.~(\ref{eq:enekin}) and~(\ref{eq:enebulk}), and the quadratic Zeeman energy density $\epsilon_{\text{zeeman}}$ is given by
\begin{align}
 \epsilon_{\text{zeeman}} = - |q| (|\psi_1 |^2 + |\psi_{-1}|^2 ).
\end{align}
Here, $q$ is the coefficient of the quadratic Zeeman term. The negative quadratic Zeeman is experimentally realized by the technique of microwave dressing~\cite{Bloch:2006}. 

To investigate the energetic stability of the vortex state, we solve the Gross-Pitaevskii equation and seek a stationary vortex state under a fixed boundary condition. We calculate the two-dimensional Gross-Pitaevskii equation for spin-1 BECs in a uniform system with the Clank-Nicolson method. The time-dependent Gross-Pitaevskii equations in the presence of a quadratic Zeeman effect are given by
\begin{subequations}
\begin{align}
i \hbar \frac{\partial \psi_1}{\partial t} &= \left( - \frac{\hbar^2 }{2 M} \nabla^2  + c_0 n  + \langle f_z \rangle  - |q| \right) \psi_1 + c_1 n\langle f_- \rangle \psi_0, \label{eq:gp1} \\
i \hbar \frac{\partial \psi_0}{\partial t} &= \left( - \frac{\hbar^2 }{2 M} \nabla^2  + c_0 n \right) \psi_0 + c_1 n ( \langle f_+  \rangle \psi_1+ \langle f_-  \rangle \psi_{-1}), \label{eq:gp0}   \\
i \hbar \frac{\partial \psi_{-1}}{\partial t} &= \left( - \frac{\hbar^2 }{2 M} \nabla^2  + c_0 n  - \langle f_z \rangle  - |q| \right) \psi_{-1} + c_1 n\langle f_+ \rangle \psi_0,  \label{eq:gp-1}
\end{align}
\end{subequations}
where $ \langle f_z \rangle$, $\langle f_+ \rangle$, and $ \langle f_- \rangle$ are given by
\begin{subequations}
\begin{align}
 &\langle f_z \rangle := \sum_{mn=-1}^1 \psi_{m}^{\ast} (f_z)_{mn} \psi_n, \\
 &\langle f_{\pm} \rangle := \frac{1}{\sqrt{2}} \sum_{mn=-1}^1 \psi_{m}^{\ast} (f_x\pm i f_y)_{mn} \psi_n,
\end{align}
\end{subequations}
respectively.
\begin{figure}[tbp]
\centering
\includegraphics[width=9cm]{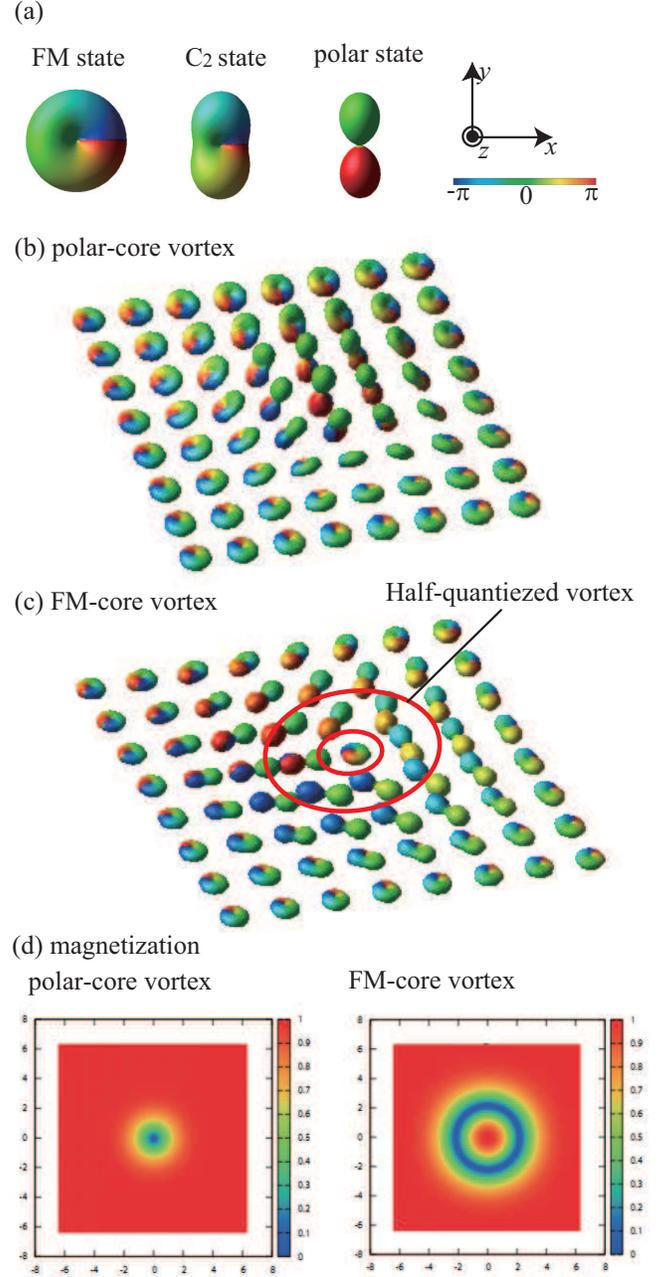}
\caption{(Color) (a) Plots of $|\psi (\theta, \phi)|$ for the FM state, the $C_2$ state, and the polar state, where $\psi(\theta, \phi) = \sum_{m=-1}^1 Y_{1 m} (\theta, \phi) \psi_m$. Here, $Y_{1m}$ is a rank-$1$ spherical-harmonic function and colors represent the phase of $\psi (\theta , \phi)$. (b) and (c) Vortex-core structures of the polar-core vortex ($|q|/|c_1|=0$) and the FM-core vortex ($|q|/|c_1| = 0.4$) by the spherical-harmonic representation. We assume $c_0/|c_1| = 100$. The polar-core vortex is consistent with the core structure in Eq.~(\ref{thm2-1}), whereas the FM-core vortex corresponds to the core structure in Eq.~(\ref{thm2-2}). In the intermediate region, there exists a half-quantum vortex along a ring of the polar state. (d) Amplitude of the magnetization $|\bm{F}|$ in the two-dimensional system. The vertical and horizontal axes show the scaled dimensionless $y$ and $x$ axes, respectively. The red and blue regions represent the FM state and the polar state, respectively. The other colors show the $C_2$ state.} \label{fig:num}
\end{figure}
We seek stable solutions by an imaginary time propagation with the fixed boundary condition. We assume $c_0 \gg |c_1|$. Therefore, the maximally enlarged OPM is $S^5$, which is consistent with our classification method discussed above. 

 For the FM state $(c_1 < 0)$, we choose the boundary condition as 
\begin{align}
\bm{\psi}_{\rm F} =\sqrt{n} (e^{i \varphi}, 0, 0)^T, \label{eq:bs-1}
\end{align} 
where $\varphi$ is the azimuthal angle in the two-dimensional space. 
The numerical results are shown in Fig~\ref{fig:num}, where we show stable solutions for $q=0$ and $q < q_{\rm c}$, respectively. Here, $q_{\rm c}$ is a transition point from the polar-core vortex to the FM-core vortex. At $q=0$, the vortex core is filled with the polar state as shown in Fig.~\ref{fig:num} (b). The vortex-core state is given by $\bm{\psi}_{\rm P} = (0,\sqrt{n},0)^T$. Since the intermediate region is covered by the $C_2 $ state, this is the polar-core vortex whose vortex-core structure corresponds to Eq.~(\ref{thm2-1}) and the order of strata is given by Eq.~(\ref{eq:fp-3}). The polar-core vortex has been realized experimentally~\cite{Sadler:2006}, and our classification method is consistent with both the experimental observation and the numerical simulation.  

On the other hand, at $q < q_{\rm c}$, the vortex core is filled with the FM state. As shown in Fig.~\ref{fig:num} (c), we have a ring of the polar state in the intermediate region. Because the component $\psi_0$ is negligible for $q < q_{\rm c}$ due to $E_{\text{Zeeman}} \propto |q| |\psi_0|^2 $. Hence, the core state becomes $\bm{\psi}_{\rm F} = (0,0,\sqrt{n})^T$ and the polar state on the ring is given by
\begin{align}
 \bm{\psi}_{\rm P }= \sqrt{\frac{n}{2}}(e^{i \varphi},0,1)^T. \label{eq:pring}
\end{align}
Here, Eq.~(\ref{eq:pring}) represents the half-quantum vortex.
Therefore, the core structure corresponds to Eq. (\ref{thm2-2}) and the corresponding order of strata is given by Eq.~(\ref{eq:fpf-3}) with $n_p = 1$. Since the vortex core has nontrivial winding, it is an excited state of the core.
 
  Next, in the polar state $(c_1 > 0)$, we choose a boundary condition as the half-quantum vortex in Eq.~(\ref{eq:pring}).  For $q \le 0$, the vortex-core is always filled with the FM state $\bm{\psi_{\rm F}} = (0,0,\sqrt{n})^T$. The order parameter in the intermediate region is always the $C_2$ state. Hence, the vortex-core structure corresponds to Eq.~(\ref{thm3-1}) and the order of strata is given by Eq.~(\ref{eq:pf-3}). This vortex-core structure has been discussed numerically in Refs.~\cite{Isoshima:2001} and \cite{Ruostekoski:2003}. The excited core state is unstable at least in the presence of a negative quadratic Zeeman effect. 
  
\section{Summary and Discussion}
\label{sec:sum}
 In this paper, we have developed a method to classify vortex-core structures. Our classification generalizes the previous works~\cite{Memin:1978,Lyuksyutov:1978} so as to identify the state that fills the vortex core when the vortex is nonsingular. Our classification method is formulated as follows: first, we define the maximally enlarged OPM, which is determined by a constant particle density, and decompose this  maximally enlarged OPM into strata. By defining the map between the strata, we have characterized the vortex-core structure. 

We have applied this method to the spin-1 BEC, and as a result, we have found that the vortex-core structure is made up of several states which are aligned  in a concentric fashion and, as a whole, becomes an excited state of the core. Each state of the concentric ring has a local winding number which constitutes a quantum number characterizing the vortex-core structure. 

For the case of the FM state, the vortex-core structure is formed from the concentric pattern of the FM state, the $C_2$ state, and the polar state. Under the minimal concentric pattern, the vortex-core structure is classified by an integer, which represents a winding number defined along a ring of the polar state in the core. 
For the case of the polar state, the vortex-core structure is also composed of the FM state, the $C_2$ state, and the polar state. There exists one nontrivial vortex-core excitation. 
We have numerically examined the energetic stability of the vortex-core structure that has a nontrivial winding number and found that it can be stabilized in the FM state under the negative quadratic Zeeman effect.  

In this paper, we have focused on the spin-1 BEC; however, our method can also be applied to the case of higher spins. These applications will be discussed elsewhere. We comment here on the role of linear Zeeman shift. Since the cold-atom systems always conserve the total spin, the linear Zeeman shift merely induces the Larmor precession, which can be eliminated by moving onto the rotating frame of reference in spin space. However, if we consider the case in which the total spin is not conserved, we cannot ignore the linear Zeeman shift. In this case, the direction of the magnetization becomes parallel to the magnetic field, and therefore the order parameter is described by a single component. Hence, the maximally enlarged OPM is $S^1$. Since $\pi_1 (S^1) \cong \mathbb{Z}$, the vortex core is singular.       

 Finally, we discuss yet another application of our classification method to the stability of a vortex under the transition between two ordered phases. We can easily modify our scheme from the classification of the  vortex core to the stability of a vortex under the transition by replacing distance $r$ with time $t$. If we apply our method to the stability of a vortex under the transition, there exists another deformation process because the initial state need not have the winding number. For example, when the initial state is the FM state and the final state is the polar state, we run though the $C_2$ state at an intermediate stage. It is possible to change to the $C_2$ state with a nontrivial winding number $n_p$ even when the FM state is uniform.
 
 {\it Note added in proof} Recently, a related work by J.~Lovegrove, {\it et al.} \cite{Lovegrove:2012} appeared which discuss as energetically stable vortex state in a spin-1 BEC.
 
\section*{ACKNOWLEDGMENTS}
The authors would like to thank D.~A.~Takahashi and M.~Kobayashi for useful discussions. This work was supported by Grants-in-Aid for Scientific Research (Kakenhi Nos. 22340114, 22103005, 22740265, and 23740198), a Grants-in-Aid for Scientific Research on Innovative Areas ``Topological Quantum Phenomena" (Nos. 22103005 and 23103515), a Global COE Program ``The Physical Science Frontier", and the Photon Frontier Netowork Program of MEXT of Japan. S.K. acknowledges support from JSPS (Grant No.228338).  Y.K. acknowledges the financial support from Inoue foundation for science.     

\appendix
\section{CALCULATION OF EQ.~(\ref{thm:op})}
\label{sec:appA}
Let us describe the calculation of Eq.~(\ref{thm:op}). A spin-1 parameter is given by 
\begin{align}
 \bm{\psi} = (\psi_1 ,\psi_0 ,\psi_{-1})^T \in \mathbb{C}^3,
\end{align}
where $\psi_m \; (m=1,0,-1)$ are described by
\begin{subequations}
\begin{align}
 \psi_1 &= x_1 + i x_2, \\
 \psi_0 &= x_3 +i x_4, \\
 \psi_{-1} & = x_5 + ix_6. \label{app:1-1}
\end{align}
\end{subequations}
Here, $x_i \in \mathbb{R}\; (i =1,2,3,4,5,6)$. To derive Eq. (\ref{thm:op}), we operate $e^{i f_y \beta} e^{i f_z \alpha}$ on $\bm{\psi}$: 
\begin{align}
& e^{i f_y \beta} e^{i f_z \alpha} \bm{\psi} \notag \\ &=  \begin{pmatrix} e^{i \alpha} \cos^2 \frac{\beta}{2} & \frac{\sin \beta}{\sqrt{2}} & e^{- i \alpha} \sin^2 \frac{\beta}{2} \\ - e^{ i \alpha} \frac{\sin \beta}{\sqrt{2}} & \cos \beta & e^{- i \alpha} \frac{\sin \beta}{\sqrt{2}}  \\ e^{i \alpha} \sin^2 \frac{\beta}{2} &- \frac{\sin \beta}{\sqrt{2}} & e^{-i \alpha} \cos^2 \frac{\beta}{2} \end{pmatrix} \left( \begin{array}{@{\,} c @{\,}} x_1 + i x_2 \\ x_3 +i x_4 \\ x_5 + ix_6 \end{array} \right). \label{app:1-2}
\end{align}
We determine $\alpha$ and $\beta$ so that the $m=0$ component vanishes. Hence, we obtain $\alpha$ and $\beta$ for $x_3 \neq 0 $ and $x_4 \neq 0$ as 
\begin{widetext}
\begin{subequations}
\begin{align}
 \alpha &= \cos^{-1} \left[ \frac{x_4 (x_2 + x_6) + x_3 (x_1 + x_5)}{\sqrt{\{ x_4 (x_5 - x_1 ) - x_3 (x_6 - x_2)\}^2 + \{ x_4 (x_2 + x_6) +x_3 (x_1 + x_6)\}^2 }}\right], \label{app:1-3a}\\
  \beta &= \cot^{-1} \left[- \frac{1}{\sqrt{2}}\frac{x_5^2 - x_1^2 +x_6^2 - x_2^2}{\sqrt{\{ x_4 (x_5 - x_1 ) - x_3 (x_6 - x_2)\}^2 + \{ x_4 (x_2 + x_6) +x_3 (x_1 + x_6)\}^2 }} \right]. \label{app:1-3b}
 \end{align}
\end{subequations}
\end{widetext}
 Therefore, the remaining components are described by $x_1'$, $x_2'$, $x_5'$, and $x_6'$  
 \begin{align}
  e^{i f_y \beta} e^{i f_z \alpha} \bm{\psi} =  \left( \begin{array}{@{\,} c @{\,}} x'_1 + i x'_2 \\ 0 \\ x'_5 + ix'_6 \end{array} \right), \label{app:1-4}
 \end{align}
 Next, we operate $e^{-i \phi} e^{i f_z \gamma}$ on Eq.~(\ref{app:1-4}) and we give $\phi$ and $\gamma$ in order for the remaining components to be real as
\begin{subequations}
\begin{align}
\gamma &= \frac{\theta_{1} - \theta_{-1}}{2}, \label{app:1-5a} \\
 \phi &= - \frac{\theta_1 + \theta_{-1}}{2}, \label{app:1-5b}
\end{align}
\end{subequations}
where $\theta_{1}$ and $\theta_{-1}$ are defined by
\begin{subequations}
\begin{align}
 \theta_1 &= \tan^{-1} \left( \frac{x'_2}{x'_1} \right), \\
 \theta_{-1 } &= \tan^{-1} \left( \frac{x'_6}{x'_5} \right). \label{app:1-6}
\end{align} 
\end{subequations}
Giving $\alpha$, $\beta$, $\gamma$, and $\phi$ as Eqs.~(\ref{app:1-3a}), (\ref{app:1-3b}), (\ref{app:1-5a}) and~(\ref{app:1-5b}), the remaining components become $\sqrt{{x'_1}^2 +{x'_2}^2}$ and $\sqrt{{x'_5}^2 +{x'_6}^2}$. Using Eqs.~(\ref{eq:dos}) and~(\ref{eq:lmag}), the order parameter is given by using $n$ and $|\bm{F}|$ as
\begin{align}
 e^{-i \phi} e^{i f_z \gamma} e^{i f_y \beta} e^{i f_z \alpha} \bm{\psi} = \sqrt{n} \left( \begin{array}{@{\,} c @{\,}} \sqrt{\frac{1 + | \bm{F} | }{2}} \\ 0 \\ \sqrt{\frac{1 - | \bm{F} | }{2}}  \end{array} \right).
 \end{align}
Thus, we obtain Eq.~(\ref{thm:op}). 

\section{CALCULATION OF $\tilde{H}_{\rm F}$, $\tilde{H}_{\rm P}$, AND $\tilde{H}_{C_2}$}
\label{sec:appB}
 In this section, we give the calculation of the lifting isotropy group for each $G$-orbit. Each isotropy group is given by Eqs.~(\ref{eq:Hf}),~(\ref{eq:Hp}), and ~(\ref{eq:Hc2}). The universal covering space of $G$ is given by Eq.~(\ref{eq:liftG}). 
To calculate $\tilde{H}_q$, we define two projection maps $p_1$ and $p_2$.
First, the map $p_1$ is defined by the map from $\tilde{G}$ to $G $ such that
\begin{align}
 p_1 :  \ \ \tilde{G} \ \ &\to \ \ G , \notag \\
  (x , e^{-i \frac{\sigma_{\nu}}{2} \theta}) &\mapsto (e^{i x}, e^{-i  f_{\nu} \theta}) \; ^{\forall}x \in \mathbb{R}, ^{\forall} \theta \in [0, 4 \pi), \label{app:2-1} 
\end{align}
where $\nu= x,y,z$, $\sigma_{\nu}$ is a Pauli matrix, and $p_1$ is a surjective and homomorphic map. Hence, since $e^{i 2 \pi n} = 1 $ and $e^{i f_{\nu} 2 \pi} = \bm{1}_{3}$, we can obtain a kernel of $p_1$ as
\begin{align}
\ker p_1 = \{ (2 \pi n , \bm{1}_{2}) , (2 \pi n , - \bm{1}_{2}) | n \in \mathbb{Z}\}, \label{app:2-2}
\end{align}
Second, $p_2$ is defined by the map from $G$ to $G/H_q$ such that
\begin{align}
 p_2 : \ \ G \ \ &\to \ \ G/H_q, \notag \\
 (e^{i \phi}, e^{-i f_{\nu} \alpha}) &\mapsto [(e^{i \phi}, e^{-i f_{\nu} \alpha})]_{q} \; ^{\forall}\phi , \alpha \in [0, 2 \pi), \label{app:2-3}
\end{align}    
where $[\cdots ]_q$ represents an equivalence class defined by
\begin{align}
 g g'^{-1} \in H_q \;  \Leftrightarrow   \; g \sim g' \; \text{for any } g, g' \in G  . \label{app:2-4}
 \end{align}
 Here, $p_2$ is also a surjective and homomorphic map. 
 By definition (\ref{app:2-2}), we obtain the kernel of $p_2$ by
 \begin{align}
  \ker p_2 = H_q. \label{app:2-5}
 \end{align}
 By using $p_1$ and $p_2$, we make up the map from $\tilde{G} $ to $G/H_q$
 \begin{align}
  p_2 \circ p_1 : \tilde{G} \to G/H_q.  \label{app:2-6}
 \end{align}
Thus, $\tilde{H}_q$ is rewritten as
\begin{align}
 \tilde{H}_q = \ker p_2 \circ p_1. \label{app:2-7}
\end{align} 
To calculate $\tilde{H}_q$, we define the inclusion map from $G$ to $\tilde{G}$ as follows:
\begin{align*}
  & \quad \vdots \\
\iota_{(- 2 \pi , \bm{1}_2)} &: (e^{i \phi}, e^{-i f_{\nu} \gamma} ) \mapsto (\phi - 2 \pi , e^{-i \frac{\sigma_{\nu}}{2} \gamma})   \ \ \phi ,\; \gamma \in [ 0, 2\pi ), \\
 \iota_{(0,\bm{1}_2)} &: (e^{i \phi}, e^{-i f_{\nu} \gamma} ) \mapsto (\phi , e^{-i \frac{\sigma_{\nu}}{2} \gamma}), \\
  \iota_{(2 \pi ,\bm{1}_2)} &: (e^{i \phi}, e^{-i f_{\nu} \gamma} ) \mapsto (\phi+2 \pi  , e^{-i \frac{\sigma_{\nu}}{2} \gamma}), \\
  & \quad \vdots \\
\end{align*}
Then, we can give the lifted isotropy group $\tilde{H}_q$ by using $\iota_s \; (s \in \ker p_1)$ as
\begin{align} 
 \tilde{H}_q &= p^{-1}_1 (H_q) \\
              & = \bigcup_{n \in \mathbb{Z}} \iota_{(2 \pi n , \bm{1}_2) } (h) \cup \bigcup_{n \in \mathbb{Z}} \iota_{ (2 \pi n, - \bm{1}_2 ) } (h) , \ \ ^{\forall} h \in H_q . \label{app:2-8}
\end{align}
We first calculate $\tilde{H}_{\rm F}$. Since $H_{\rm F}$ is given by Eq. (\ref{eq:Hf}), $\tilde{H}_{\rm F}$ is obtained as
\begin{widetext}
\begin{align}
 \tilde{H}_{\rm F} &= \bigcup_{n \in \mathbb{Z}} \iota_{(2 \pi n , \bm{1}_2) } ((e^{i \theta }, e^{- i f_z \theta})) \cup \bigcup_{n \in \mathbb{Z}} \iota_{ (2 \pi n, - \bm{1}_2 ) } ((e^{i \theta }, e^{- i f_z \theta})) , \ \ \theta \in [0,2 \pi ), \notag \\
 &= \cdots \cup (\theta - 2 \pi , e^{- i \frac{\sigma_z}{2}} \theta) \cup (\theta ,e^{-i\frac{\sigma_z}{2} \theta }) \cup (\theta + 2 \pi ,e^{-i\frac{\sigma_z}{2} \theta }) \cup \cdots \cup (\theta  + 2 \pi n,e^{-i\frac{\sigma_z}{2} \theta }) \cup \cdots, \notag \\ 
 & \cdots \cup (\theta - 2 \pi , - e^{- i \frac{\sigma_z}{2}} \theta ) \cup (\theta , -e^{-i\frac{\sigma_z}{2} \theta }) \cup (\theta + 2 \pi , -e^{-i\frac{\sigma_z}{2} \theta }) \cup \cdots \cup (\theta  + 2 \pi n, -e^{-i\frac{\sigma_z}{2} \theta }) \cup \cdots , \notag \\
&= \cdots \cup (\theta - 2 \pi ,  e^{-i \frac{\sigma_z}{2} (\theta - 2 \pi)}) \cup  (\theta ,  e^{-i \frac{\sigma_z}{2} \theta}) \cup  (\theta + 2 \pi , e^{-i \frac{\sigma_z}{2} (\theta + 2 \pi)}) \cup \cdots, \notag \\
 & \cdots \cup (\theta - 2 \pi , - e^{-i \frac{\sigma_z}{2} (\theta - 2\pi )}) \cup  (\theta  ,  - e^{-i \frac{\sigma_z}{2} \theta}) \cup  (\theta + 2 \pi ,  - e^{-i \frac{\sigma_z}{2} (\theta +2\pi )}) \cup \cdots , \notag \\
 &= (x , e^{i \frac{\sigma_z}{2}x}) \cup (x , - e^{i \frac{\sigma_z}{2}x}),  \label{app:2-9}
\end{align}
\end{widetext}
where $x \in \mathbb{R}$ and $e^{ \pm i\frac{\sigma_i}{2} 2 \pi} = - \bm{1}_{2}$.
Next, we calculate $\tilde{H}_{\rm P}$. Since $H_{\rm P}$ is given by Eq. (\ref{eq:Hp}), $\tilde{H}_{\rm P}$ is given by 
\begin{align}
\tilde{H}_{\rm P} &= \bigcup_{n \in \mathbb{Z}} \iota_{(2 \pi n , \pm \bm{1}_2) } ((1,  e^{-i f_y \beta }))  \\
                   &\qquad \qquad \cup \bigcup_{n \in \mathbb{Z}} \iota_{ (2 \pi n, \pm \bm{1}_2 ) } ((e^{i \pi} , e^{-i f_y \beta} e^{ -i f_z \pi})),  \notag  \\
 & =   (2 \pi n , e^{-i \frac{\sigma_y}{2} \theta}) \cup \left( (2n+1 )\pi  , e^{-i \frac{\sigma_y}{2} \theta} e^{-i\frac{\sigma_z}{2} \pi }\right),
\end{align}
where $n \in \mathbb{Z}$ and $\theta \in [0, 4 \pi )$. 
Finally, we calculate $\tilde{H}_{C_2}$. Since $H_{C_2}$ is given by Eq. (\ref{eq:Hc2}), $\tilde{H}_{C_2}$ is given by
\begin{align}
  \tilde{H}_{C_2} &= \bigcup_{n \in \mathbb{Z}} \iota_{(2 \pi n ,  \pm \bm{1}_2) } ((1, \bm{1}_{3})) \cup \bigcup_{n \in \mathbb{Z}} \iota_{ (2 \pi n, \pm \bm{1}_2 ) } ((e^{i \pi} , e^{ -i f_z \pi})), \notag \\
  & = (2\pi n , \pm \bm{1}_{2}) \cup \left( (2n+1) \pi  , \pm e^{- i \frac{\sigma_z}{2} \pi }\right) .
\end{align}
where $n \in \mathbb{Z}$.
Therefore, $\tilde{H}_{C_2} $ is composed of discrete elements.


\begin{thebibliography}{999}
\bibitem{text1}
The Mermin-Ho and Anderson-Thouless textures are not nonsingular vortices according to our definition because these textures do not leave the OPM. They can be characterized by the higher homotopy group rather than the fundamental group.
\bibitem{text2}
If $X_i \cap \tilde{H}_{ \langle \psi' \rangle}  = \emptyset $, Eq. (\ref{map:chi-1}) does not have a mapped element. We do not discuss such a case in this paper.
\end{thebibliography}
\end{document}